\documentclass[a4paper,fleqn,usenatbib]{mnras}

\usepackage{newtxtext,newtxmath}

\usepackage[T1]{fontenc}
\usepackage{ae,aecompl}

\usepackage{graphicx}	
\usepackage{epstopdf}   
\usepackage{amsmath}	

\newcommand{\kms}{\,km\,s$^{-1}$}	
\newcommand\bca{\,$\beta$\,Cyg\,A}	
\newcommand\bcb{\,$\beta$\,Cyg\,B}	
\newcommand\gaia{\textit{Gaia}}
\newcommand\hip{\textsc{Hipparcos}}

\newcommand\gdrtwo{\gaia~DR2}
\newcommand\gdrthree{\gaia~DR3}
\newcommand{\masyr}{\,mas\,yr$^{-1}$}	

\title[The Albireo system]{A celestial matryoshka: Dynamical and spectroscopic analysis of the Albireo system }

\author[R. Drimmel et al.]{
Ronald Drimmel,$^{1}$\thanks{E-mail: ronald.drimmel@inaf.it (RD)}
Alessandro Sozzetti,$^{1}$
Klaus-Peter Schr\"{o}der,$^{2,3}$
Ulrich Bastian,$^{4}$
\newauthor
Matteo Pinamonti,$^{1}$
Dennis Jack$^{2}$
and Missael A. Hern\'{a}ndez Huerta$^{2}$
\newauthor
\\
$^{1}$Istituto Nazionale di Astrofisica, Osservatorio Astrofisico di Torino, Via Osservatorio 20, 10024 Pino Torinese (TO), Italy\\
$^{2}$Departamento de Astronom\'ia, Universidad de Guanajuato, Callejón de Jalisco S/N, 36023 Guanajuato, GTO, Mexico\\
$^{3}$Sterrewacht Leiden, University of Leiden,P.O. Box 9513, NL-2300 RA, Leiden, The Netherlands\\
$^{4}$Zentrum f\"ur Astronomie (Center for Astronomy), Heidelberg University, M\"onchhofstr. 14, D-69120 Heidelberg, Germany
}

\date{Accepted XXX. Received YYY; in original form ZZZ}

\pubyear{2019}

\begin{document}
\label{firstpage}
\pagerange{\pageref{firstpage}--\pageref{lastpage}}
\maketitle

\begin{abstract}
We present a spectroscopic characterisation and a new orbital solution for the 
binary system \bca a/Ac (MCA 55), the primary component (\bca) of the 
well-known wide double star Albireo.  By matching evolutionary 
tracks to the physical parameters of all three Albireo stars (\bca a, Ac and B) 
as obtained from a spectroscopic analysis of TIGRE and IUE spectra, we confirm that they are likely coeval. Our final orbit solution is based on radial-velocity measurements taken over a baseline exceeding 120 years, combined with relative astrometry from speckle interferometric observations and the absolute astrometry from the \hip\ and \gaia\ missions. Our final orbit solution has a period of 121.65$^{+3.34}_{-2.90}$ years with an eccentricity of 0.20$^{+0.01}_{-0.02}$. Thanks to the inclusion
of the absolute astrometry, we find a mass ratio of $q = 1.25^{+0.19}_{-0.17}$, and a total mass of $9.47_{-3.24}^{+5.88}$\, M$_\odot$, indicating that the secondary (Ac) is the more massive of the pair. These results strongly suggest the presence of a fourth, unseen, member of the Albireo system. Given the current photometric data it is likely that \bca\ is itself a hierarchical triple. 
We also derive the systemic proper motion, line-of-sight velocity, and an orbital parallax of the \bca\ system, allowing us to quantitatively assess the hypothesis that Albireo A and B form a physically bound and genealogically connected system. Finally, we 
find four potential members of a common proper motion group with Albireo, though none 
anywhere as close by as the Albireo components A to B.
\end{abstract}




\section{Introduction}



Albireo ($\beta$\,Cyg) is a well-known naked-eye object that is a popular target for amateur astronomers, as it is easily resolved into a beautiful pair of stars with starkly contrasting colors. The binary nature of \bca B has been the subject of a long-standing debate which many hoped \gaia\ astrometry would finally resolve, but the \gdrtwo\ parallaxes for both stars are too uncertain to settle the question and, in any case, cannot be considered reliable given the brightness of the two components \citep{Drimmel2019}. Meanwhile, the \gdrtwo\ proper motions for \bca\ and B are completely different, suggesting that the pair is only a casual double. However, as pointed out by \citet{bastian18}, the primary \bca\ is itself an unresolved binary for \gaia (at least up to \gdrtwo), and the orbital motion of the brighter component does significantly contribute to the measured proper motion of \bca, given the short baseline of the observations contributing to the \gdrtwo\ astrometry, so that \bca B may still be a bound triple system.  Only once the orbit of the close pair \bca\ (Aa,Ac) is well determined can one hope to "correct" the measured proper motions of \bca\ to arrive at the systemic motion of \bca\ and, in turn, its possible physical connection to \bcb.

The binary nature of \bca\ has long been known: Given its brightness, \bca\ was an obvious target for the first spectroscopic observations, and it was soon recognized as having a composite spectrum \citep{Maury1897, Clerke1899} with a dominant "cool" stellar component but clear evidence of a fainter "hot" component. More recent studies have determined the spectra as being a K3II giant with a B9V companion \citep{Markowitz69, Parsons1998}. 
As one of the first recognized bright unresolved doubles, \bca\ became a target of interest in many of the earliest spectroscopic observing programs. But due to the long orbital period and small radial-velocity amplitude, no orbital solution based on its radial velocities has been published to date. 

Previous to this work, available orbit solutions of \bca\ have been based solely on speckle observations measuring the relative astrometry of Aa/Ac, taken on a semiregular basis since 1976 by different observing programs. Preliminary orbit solutions were first presented by \citet{Hartkopf1999} ($P=96.84$ yrs), followed by \citet{Scardia2007} ($P=213.859$ yrs). 

Using the orbit solution by \citet{Scardia2007}, together with the extant astrometry from \hip\ and \gaia, \citet{bastian18} argued that the inferred systemic proper motion of the \bca\ system would be inconsistent with the proper motion of \bcb, making \bca B a casual (optical) double. But their analysis also implied an implausibly small mass for the \bca\ primary.  However, more recently \citet{Jack2018} confirmed that the \bca\ primary (Aa) is a typical red K giant, based on high-resolution spectroscopic measurements. These reveal that its surface gravity is consistent with having a mass of 5 $M_\odot$. Thus the \citet{Scardia2007} orbit was falsified, and the possibility that \bca B is a physical triple system cannot be excluded.

Later \citet{Roberts2018} were able to present a formal orbit solution ($P = 68.6 \pm 5.8$ yrs), thanks mainly to additional speckle observations accumulated by others. However, this later orbit led to an non-physical total mass of the system of 85\,$M_\odot$, and is in stark contrast to the \gaia\ DR2 proper motion of \bca\ (see \citet{bastian18}). Most recently a new orbit solution has been published by \citet{IAU198} using new speckle data, with a period of 120 years. However, its uncertainties are such that even this one must still be considered a preliminary solution (Scardia, private communication). 

After a presentation of the data in Section \ref{sec:data}, in Section\,\ref{sec:spectro} we present a new careful spectroscopic analysis and derive physical parameters of the two stars of \bca\ as well as of \bcb, and estimate their ages under the assumption of a common distance.  A first, well-constrained orbit solution for \bca a/Ac is presented in Section\,\ref{sec:orbit}, based on both speckle and radial velocity measurements, with additional constraints from the \hip\ and \gaia\ astrometry that allow us to constrain the total mass, distance and systemic velocities of the system. Using these, we evaluate the relationship between \bca\ and B in Sections\,\ref{sec:dynamical} and \ref{sec:kinematic}, assessing the possibility that they are a bound system with common origin, and we suggest that four stars may be additional stellar members belonging to it. Finally, we present the evidence that \bca\ contains an additional, unseen companion in Section \ref{sec:discussion}, and end with a concluding discussion.

\section{Data}
\label{sec:data}

\subsection{Radial velocity data}
\label{sec:RVdata} 

For the purpose of orbit reconstruction the most useful of the early observing programs is that of Lick Observatory, which culminated in a large catalogue of radial velocities \citep{LickCat}. This compilation contains 29 observations of \bca\ taken over a period of 28 years using two different spectrographic instruments mounted on the 36 inch refractor on Mt. Hamilton. Uncertainties on the individual measurements are not provided, so we have assigned to these a provisional (and admittedly optimistic) uncertainty of 1\kms, though we expect that the observations from the later New Mills instrument are of better quality than those from its predecessor. In any case, the uncertainties for these and the following radial velocity datasets will be checked against the standard deviation of the normalized residuals with respect to a satisfactory orbit solution. 

In the same time period as covered by the Lick Observatory program only a few observations from other observatories can be found, but over much smaller temporal baselines, and usually of lower quality.  Later radial velocity catalogues list $\beta$\,Cyg\,A as a single entry, giving its mean radial velocity (RV) derived from the compilation of the available observations from multiple observing programs over many years (e.g. \citet{GCRV}), so are not useful for orbit determination.

Following the Lick Observatory program there is a long period of more than 40 years during which RV measurements of \bca\ are few and of very limited quality, though it is worth mentioning the compilation of \citet{Hendry81}, who also contributed her own observations taken over a twelve year period. 
From this heterogeneous compilation we take only the observations made by E. Hendry herself, made with the Northwestern University 1.1m LARC telescope
\footnote{Identification of the observatory for each observation was confirmed with the assistance of E. Hendry, private communication.}. We converted her quoted {\em probable errors} to standard errors assuming that these were estimated from the standard error $\sigma$, i.e. $\gamma = 0.6745 \sigma$.
Though this data set is of inferior quality, it partially fills the large gap between Lick and modern instruments and, as discussed later, its inclusion improves the quality of resulting orbit solutions.

Fortunately, modern RV surveys would soon begin to observe \bca\ with a regular cadence and with superior instrumentation than was previously available. In particular, here we present two new datasets: 14 individual RV measurements made with CORAVEL on the 1 m Swiss telescope at the Haute-Provence Observatory made between 1981 and 1998 \citep{Famaey2005}, kindly provided by B. Famaey (private communication), followed by 12 recent RV measurements based on a 
series of high-resolution ($R\approx 20,000$) spectra with the HEROS 
spectrograph at the TIGRE telescope \citep{Schmitt2014}, using the methodology described in \citet{Mittag2018}. 

In summary, we have RV observations of the primary component \bca a spanning a baseline of more than 120 years, though with a large gap of about 40 years. The complete set of RV measurements is given in Table \ref{tab:RVs_tbl}, and shown in Figure \ref{fig:rv_model}.

\begin{table}
	\centering
	\caption{Radial velocity data and associated errors, in \kms. The epoch is in Julian Days and $N_{\rm m}$ is the number of independent measurements.}
	\label{tab:RVs_tbl}
	{\footnotesize
	\begin{tabular}{lcccc} 
		\hline
   RV &  $\sigma_{RV}$ &   epoch JD & $N_{\rm m}$ & Observatory/Instrument\\
   \hline
-25.2 & 1 & 2414421.99 & 1 & Lick/Original Mills \\
-26.2 & 1 & 2414482.79 & 1 & Lick/Original Mills \\
-24.5 & 1 & 2414785.99 & 1 & Lick/Original Mills \\ 
-25.3 & 1 & 2415239.76 & 1 & Lick/Original Mills \\
-25.8 & 1 & 2415493.02 & 1 & Lick/Original Mills \\
-24.55 & 1 & 2416687.84 & 2& Lick/New Mills \\
-25.15 & 1 & 2417747.86 & 2& Lick/New Mills \\
-26.1 & 1 & 2418906.76 & 2 & Lick/New Mills \\
-25.5 & 1 & 2419033.58 & 1 & Lick/New Mills \\
-24.5 & 1 & 2419168.93 & 2 & Lick/New Mills \\
-24.2 & 1 & 2419353.6 & 2  & Lick/New Mills \\
-23.2 & 1 & 2419570.77 & 2 & Lick/New Mills \\ 
-23.2 & 1 & 2419571.97 & 2 & Lick/New Mills \\ 
-23.8 & 1 & 2419869.01 & 1 & Lick/New Mills \\
-24.4 & 1 & 2419890.92 & 1 & Lick/New Mills \\ 
-23.5 & 1 & 2419891.92 & 1 & Lick/New Mills \\
-24.1 & 1 & 2420196.06 & 1 & Lick/New Mills \\
-23.4 & 1 & 2420197.08 & 1 & Lick/New Mills \\
-22.5 & 1 & 2420201.05 & 1 & Lick/New Mills \\
-22.8 & 1 & 2421288.06 & 1 & Lick/New Mills \\
-23.6 & 1 & 2421289.07 & 1 & Lick/New Mills \\
-22.0 & 1 & 2421360.92 & 2 & Lick/New Mills \\
-23.05 & 1 & 2421388.98 & 2& Lick/New Mills \\ 
-22.1 & 1 & 2421783.91 & 1 & Lick/New Mills \\
-21.6 & 1 & 2421876.63 & 1 & Lick/New Mills \\
-22.4 & 1 & 2422962.66 & 1 & Lick/New Mills \\
-22.4 & 1 & 2423203.99 & 1 & Lick/New Mills \\
-23.2 & 1 & 2424689.98 & 1 & Lick/New Mills \\
-22.8 & 1 & 2424706.00 & 1 & Lick/New Mills \\
-23.4 & 2.7 & 2440128.072 & 1 & NWU LARC  \\ 
-22.4 & 1.9 & 2440131.092 & 1 & NWU LARC  \\
-23.1 & 1.6 & 2442243.181 & 1 & NWU LARC  \\
-23.2 & 3.9 & 2442616.183 & 1 & NWU LARC  \\
-23.5 & 2.2 & 2442972.123 & 1 & NWU LARC  \\ 
-24.3 & 4.2 & 2444046.247 & 1 & NWU LARC  \\
-22.7 & 4.7 & 2444364.271 & 1 & NWU LARC  \\
-25.2 & 4.7 & 2444403.239 & 1 & NWU LARC  \\
-23.55 & 0.29 & 2444859.341 & 1 &   HPO/CORAVEL \\ 
-23.62 & 0.28 & 2446586.511 & 1 &   HPO/CORAVEL \\
-23.84 & 0.28 & 2446959.624 & 1 &   HPO/CORAVEL \\
-23.47 & 0.29 & 2447082.249 & 1 &   HPO/CORAVEL \\
-23.67 & 0.33 & 2447472.267 & 1 &   HPO/CORAVEL \\
-24.03 & 0.33 & 2447649.66 & 1  &   HPO/CORAVEL \\
-24.29 & 0.33 & 2447831.327 & 1 &   HPO/CORAVEL \\ 
-23.97 & 0.35 & 2448292.759 & 1 &   HPO/CORAVEL \\
-24.26 & 0.35 & 2448849.339 & 1 &   HPO/CORAVEL \\
-24.25 & 0.34 & 2449179.518 & 1 &   HPO/CORAVEL \\
-24.64 & 0.29 & 2449569.489 & 1 &   HPO/CORAVEL \\
-24.61 & 0.35 & 2450443.21 & 1 &    HPO/CORAVEL \\
-24.60 & 0.34 & 2450702.346 & 1 &   HPO/CORAVEL \\
-24.72 & 0.36 & 2451005.529 & 1 &   HPO/CORAVEL \\ 
-25.07 & 0.17 & 2458391.554 & 1 &  TIGRE/HEROS  \\
-25.29 & 0.15 & 2458446.559 & 1 &  TIGRE/HEROS \\
-25.71 & 0.12 & 2458527.019 & 1 &  TIGRE/HEROS \\
-25.78 & 0.12 & 2458528.014 & 1 &  TIGRE/HEROS \\
-25.73 & 0.12 & 2458529.013 & 1 &  TIGRE/HEROS \\
-25.79 & 0.12 & 2458577.969 & 1 &  TIGRE/HEROS \\
-25.70 & 0.11 & 2458612.927 & 1 &  TIGRE/HEROS  \\
-25.63 & 0.12 & 2458624.889 & 1 &  TIGRE/HEROS \\ 
-25.71 & 0.11 & 2458671.814 & 1 &  TIGRE/HEROS \\ 
-25.06 & 0.16 & 2458717.658 & 1 &  TIGRE/HEROS \\
-25.10 & 0.16 & 2458748.564 & 1 &  TIGRE/HEROS \\
-25.16 & 0.17 & 2458770.596 & 1 &  TIGRE/HEROS \\
\hline
	
	\end{tabular}
	}
\end{table}


\subsection{Speckle observations}
\label{sec:astrometry} 

For the speckle observations of \bca\ we use the compilation in the Fourth Catalog of Interferometric Measurements of Binary Stars. (Hereafter FCIM, described in \citet{Hartkopf2001}.)\footnote{Currently hosted at
http://www.astro.gsu.edu/wds/int4.html}, maintained by the USNO up to January of 2018. These observations span from 1976 to 2008. However, many of the observations for \bca\ in the FCIM (listed under its WDS identifier 193043.29+275734.9) are missing uncertainties. We therefore assign uncertainties for all the Center for High Angular Resolution Astronomy (CHARA) speckle observations (those indicated with technique code "Sc") as prescribed by Table 3 of \citet{Hartkopf2000}, while other missing uncertainties were recovered by consulting the original publications. Care was also taken to check that the uncertainties in the separation $\rho$ were in consistent units, as the FCIM often reports the uncertainties as quoted in the original citations, where one sometimes find relative uncertainties (i.e. $\delta \rho / \rho$) or the uncertainty in $\rho$ in arcseconds. For convenience we list the FCIM observations, with the uncertainties just mentioned, in Table \ref{tab:speckl_tbl}, where we also give additional information about the telescope and instrumentation used, in order to check for possible systematics between different telescope/instrument combinations. A literature search confirmed that additional speckle observations of \bca\ have not been published since 2010, however Marco Scardia has kindly provided his most recent observation reported in \citet{IAU198}, which is the last entry of Table \ref{tab:speckl_tbl}. In summary, we note that most of the speckle observations come from one of two observing programs, the CHARA program covering the first 20 years, and those with the Pupil Interferometry Speckle camera and COronagraph (PISCO) instrument from 1995 and onwards.

\begin{figure}
	\includegraphics[width=\columnwidth]{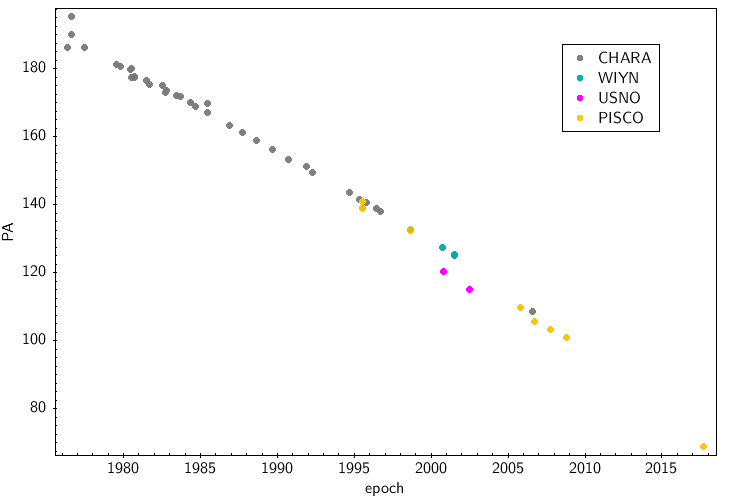}
    \caption{Position angle (PA) observations, in degrees, with respect to observing epoch.}
    \label{fig:PAplot}
\end{figure}

\begin{figure}
	\includegraphics[width=\columnwidth]{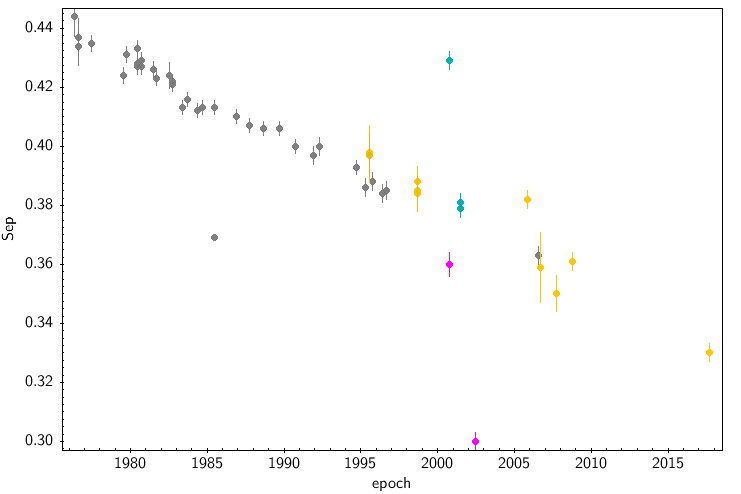}
    \caption{Separation, in arcseconds, with respect to observing epoch.}
    \label{fig:Sep_plot}
\end{figure}

Taken together, the observed position angle (PA) measures show a smooth trend, with the exception of the two USNO observations which show a clear offset, as can be seen in Figure \ref{fig:PAplot}. Meanwhile, for the measured separations (Figure \ref{fig:Sep_plot}), we note that a number of the WIYN and USNO observations show large deviations from the otherwise smooth trend. For the purpose of orbit parameter estimation we therefore exclude the two USNO observations. In addition, we exclude one observation from the CHARA program: the single observation made with the 0.6 meter Lowell refractor (epoch JD1985.4729) with no estimated uncertainties. 

\begin{table*}
	\centering
	\caption{Speckle data. Position angle (PA) and uncertainties are in degrees, separation ($\rho$) is in arcseconds, the telescope and detector columns are derived from the source references, while the codes for these and technique code (last column) are taken directly from the FCIM.}
	\label{tab:speckl_tbl}
	\begin{tabular}{lcccccccr} 
		\hline
   epoch     &   PA    &   $\sigma_{PA}$   &  $\rho$  &  $\sigma_\rho/\rho$   &   Telescope &  detector  &   reference code &  Tech code \\
   \hline
  1976.3676  &  186.2  &   0.5   &   0.444  &    0.015   &   2.1 KPNO &  phot &   McA1982b & Sc \\
  1976.6133  &  195.3  &   0.5   &   0.434  &    0.015   &   2.1 KPNO &  phot &   McA1982b & Sc \\
  1976.6217  &  190.0  &   0.5   &   0.437  &    0.015   &   2.1 KPNO &  phot &   McA1982b & Sc \\
  1977.4816  &  186.1  &   0.3   &   0.435  &    0.006   &   3.8 KPNO &  phot &   McA1979b & Sc \\
  1979.5295  &  181.2  &   0.3   &   0.424  &    0.006   &   3.8 KPNO &  phot &   McA1982d & Sc \\
  1979.7699  &  180.7  &   0.3   &   0.431  &    0.006   &   3.8 KPNO &  phot &   McA1982d & Sc \\
  1980.4795  &  179.6  &   0.3   &   0.433  &    0.006   &   3.8 KPNO &  phot &   McA1983  & Sc \\
  1980.4823  &  180.0  &   0.3   &   0.428  &    0.006   &   3.8 KPNO &  phot &   McA1983  & Sc \\
  1980.4854  &  177.5  &   0.3   &   0.427  &    0.006   &   3.8 KPNO &  phot &   McA1983  & Sc \\
  1980.7173  &  177.4  &   0.3   &   0.427  &    0.006   &   3.8 KPNO &  phot &   McA1983  & Sc \\
  1980.7255  &  177.8  &   0.3   &   0.429  &    0.006   &   3.8 KPNO &  phot &   McA1983  & Sc \\
  1981.4735  &  176.4  &   0.3   &   0.426  &    0.006   &   3.8 KPNO &  phot &   McA1984a & Sc \\
  1981.7003  &  175.3  &   0.3   &   0.423  &    0.006   &   3.8 KPNO &  phot &   McA1984a & Sc \\
  1982.5277  &  175.1  &   0.5   &   0.424  &    0.010   &   1.8 Perk &  oCDD &   Fu1997   & Sc \\
  1982.7542  &  173.0  &   0.3   &   0.422  &    0.006   &   3.8 KPNO &  oCDD &   McA1987b & Sc \\
  1982.7651  &  173.6  &   0.3   &   0.421  &    0.006   &   3.8 KPNO &  oCDD &   McA1987b & Sc \\
  1983.4175  &  172.1  &   0.3   &   0.413  &    0.006   &   3.8 KPNO &  nCCD &   McA1987b & Sc \\
  1983.7098  &  171.7  &   0.3   &   0.416  &    0.006   &   3.8 KPNO &  nCCD &   McA1987b & Sc \\
  1984.3733  &  170.0  &   0.3   &   0.412  &    0.006   &   3.8 KPNO &  nCCD &   McA1987b & Sc \\
  1984.7010  &  169.0  &   0.3   &   0.413  &    0.006   &   3.8 KPNO &  nCCD &   Hrt2000a & Sc \\
  1985.4729  &  169.8  &         &   0.369  &            &   0.6 Low  &  nCCD &   McA1987b & Sc \\
  1985.4816  &  167.2  &   0.3   &   0.413  &    0.006   &   3.8 KPNO &  nCCD &   McA1987b & Sc \\
  1986.8883  &  163.2  &   0.3   &   0.410  &    0.006   &   3.8 KPNO &  nCCD &   McA1989  & Sc \\
  1987.7618  &  161.2  &   0.3   &   0.407  &    0.006   &   3.8 KPNO &  nCCD &   McA1989  & Sc \\
  1988.6575  &  159.0  &   0.3   &   0.406  &    0.006   &   3.8 KPNO &  nCCD &   McA1990  & Sc \\
  1989.7112  &  156.3  &   0.3   &   0.406  &    0.006   &   3.8 KPNO &  nCCD &   Hrt2000a & Sc \\
  1990.7434  &  153.3  &   0.3   &   0.400  &    0.006   &   3.8 KPNO &  nCCD &   Hrt1992b & Sc \\
  1991.8959  &  151.2  &   0.3   &   0.397  &    0.008   &   3.8 KPNO &  nCCD &   Hrt1994  & Sc \\
  1992.3105  &  149.5  &   0.3   &   0.400  &    0.008   &   3.8 KPNO &  nCCD &   Hrt1994  & Sc \\
  1994.7080  &  143.6  &   0.3   &   0.393  &    0.006   &   3.8 KPNO &  nCCD &   Hrt2000a & Sc \\
  1995.3141  &  141.5  &   0.4   &   0.386  &    0.008   &   2.5 MWO &   nCCD &   Hrt1997  & Sc \\
  1995.556   &  139.0  &   0.8   &   0.398  &    0.023   &   2.0 PdM &  P-CAR &   Pru2002b & S  \\
  1995.556   &  140.9  &   1.0   &   0.397  &    0.023   &   2.0 PdM &  P-CAR &   Pru2002b & S  \\
  1995.559   &  139.3  &   1.0   &   0.397  &    0.018   &   2.0 PdM &  P-CAR &   Pru2002b & S  \\
  1995.7620  &  140.7  &   0.4   &   0.388  &    0.008   &   2.5 MWO &   nCCD &   Hrt1997  & Sc \\
  1996.4227  &  139.0  &   0.4   &   0.384  &    0.008   &   2.5 MWO &   nCCD &   Hrt2000a & Sc \\
  1996.6984  &  138.1  &   0.4   &   0.385  &    0.008   &   2.5 MWO &   nCCD &   Hrt2000a & Sc \\
  1998.657   &  132.3  &   0.9   &   0.385  &    0.018   &   2.0 PdM &  P-CCD &   Pru2002b & S  \\
  1998.657   &  132.7  &   0.5   &   0.384  &    0.016   &   2.0 PdM &  P-CCD &   Pru2002b & S  \\
  1998.657   &  132.7  &   0.5   &   0.388  &    0.013   &   2.0 PdM &  P-CCD &   Pru2002b & S  \\
  2000.7614  &  127.3  &   1.0   &   0.429  &    0.007   &   3.5 WIYN &   CCD &   Hor2002a & S  \\
  2000.7854  &  120.3  &   0.5   &   0.36   &    0.011   &   0.7 USNO &  iCCD &   WSI2001b & Su \\
  2001.4930  &  125.3  &   0.6   &   0.379  &    0.008   &   3.5 WIYN & RYTSI &   Hor2008  & S  \\
  2001.4930  &  125.0  &   0.6   &   0.381  &    0.008   &   3.5 WIYN & RYTSI &   Hor2008  & S  \\
  2002.473   &  115.1  &   0.6   &   0.30   &    0.010   &   0.7 USNO &  iCCD &   WSI2004a & Su \\
  2005.820   &  109.7  &   0.3   &   0.382  &    0.008   &   1.0 Zeiss & PISCO &  Sca2008a & S  \\
  2006.5723  &  108.6  &   0.4   &   0.363  &    0.008   &   2.5 MWO &   nCCD &   Hrt2009  & Sc \\
  2006.721   &  105.5  &   1.3   &   0.359  &    0.033   &   1.0 Zeiss & PISCO &  Sca2009a & S  \\
  2007.770   &  103.3  &   0.4   &   0.350  &    0.017   &   1.0 Zeiss & PISCO &  Sca2010c & S  \\
  2008.802   &  100.8  &   1.4   &   0.361  &    0.008   &   1.0 Zeiss & PISCO &  Pru2010  & S  \\
  2017.6873  &   68.9  &   0.3   &   0.330  &    0.009   &   1.0 CERGA & PISCO &  Sca2019  & S  \\
        \hline	
	\end{tabular}
\end{table*}

\subsection{Absolute astrometry}
\label{sec:absolute_ast}

As mentioned in the Introduction, given its brightness, no reliable parallax for \bca\ is available. However, for the purpose of estimating physical parameters 
in the following sections, we consider the hypothesis that \bca\ is at the same distance as \bcb. For this star we find that the \hip\ \citep{Perryman1997, NewHipp2007} and \gdrtwo\ 
parallaxes are consistent, with a weighted mean of $8.33 \pm 0.13$ mas, from which we adopt 120 pc as its nominal distance. 

\begin{table*}
	\centering
\caption{Astrometry data with uncertainties. The third column reports the catalogue origin of the proper motions, with T = Tycho-2, H = New \hip , G = \gdrtwo, H' and G' = cross-calibrated values from the \citet{Brandt2019} catalogue, B = proper motions from H' and G' positions. Proper motion units are all \masyr , and parallax units in mas. } 
	\label{tab:astrom_tbl}
	\begin{tabular}{lcccccccccc} 
		\hline
   object     &   sourceid   & catalogue &   $\mu_\alpha$   &  $\sigma_{\mu_\alpha}$  &   $\mu_\delta$   &  $\sigma_{\mu_\delta}$  & $\varpi$ & $\sigma_\varpi$ \\
   \hline
  \bca\ & 2133-2964-1         & T   & $-1.5$   & 0.3    & $-1.4$    & 0.3  & & \\
        & 95947               & H   & $-7.17$  & 0.25   & $-6.15$   & 0.33 & 7.51 & 0.33\\
        & 2026116260302988160 & G   & 6.127    & 1.164  & $-15.488$ & 1.091 & 9.95 & 0.60 \\
        &                     & H'  & $-7.06$  & 0.43   &   $-5.77$ &  0.53  & &\\
        &                     & G'  &  6.126   & 2.104  & $-15.488$ &  1.972  & &\\
        &                     & B   & $-2.038$ & 0.040  & $-10.018$ &  0.042  & &\\
 \hline 
  \bcb\ & 2133-2963-1         & T   & $-0.5$   & 1.2    & $-1.9$    & 1.1 & &\\
        & 95951               & H   & $-1.9$   & 0.19   & $-1.02$   & 0.27 & 8.16 & 0.25 \\
        & 2026113339752723456 & G   & $-0.990$ & 0.261  & $-0.541$  & 0.275 & 8.38 & 0.16 \\  
        &                     & H'  & $-1.880$ &  0.412 &  $-0.805$ & 0.508  & & \\
        &                     & G'  & $-0.990$ &  0.471 &  $-0.541$ &  0.497 & & \\
        &                     & B   & $-1.044$ &  0.016 &  $-1.442$ &  0.019 & & \\

        \hline	
	\end{tabular}
\end{table*}

It is possible to use the $\sim25$ yr long temporal 
baseline provided by the \hip\ and \gaia\ position measurements, as well as their absolute proper motions, to trace the orbital motion of $\beta$ Cyg Aa due 
to $\beta$ Cyg Ac. This is a desirable addition, as \hip\ and \gaia\ astrometry provides constraints complementary to those of the RV and speckle imaging astrometry, 
allowing in particular to directly infer the mass ratio. Evidence of orbital motion effects due to a perturbing companion is obtained by measuring proper motion 
changes between appropriately cross-calibrated catalogues. This approach is usually referred to as 'proper motion difference', 'astrometric acceleration', 
or 'proper motion anomaly' technique, in short $\Delta\mu$. Past applications of the methodology include catalogues of $\Delta\mu$ binaries 
\citep{Wielen2000, Makarov2005, Tokovinin2012} produced by comparison of catalogues (e.g., \hip ) including short-term proper motions (that capture the reflex orbital 
motion of the primary) with those (e.g., Tycho-2) based on long-term observations of star positions (for which the long-term proper motion will be closer to the 
true center-of-mass motion of the system). More recently, the $\Delta\mu$ technique has been applied to detect (or place upper limits on) stellar and substellar 
companions using the \hip\ and \gdrtwo\ catalogues 
\citep{Calissendorff2018, Snellen2018, Kervella2019, Brandt2019, Dupuy2019, Feng2019, Grandjean2019, Kervella2020, Damasso2020, Derosa2020, Xuan2020}. 
In such cases, the long-term proper motion vector (assumed to be describing the barycenter tangential velocity) is determined from the difference in astrometric 
position between the two catalogues divided by the corresponding $\sim25$-yr time baseline. By subtracting this long-term proper motion from the quasi-instantaneous proper motions 
of the two catalogues one obtains a pair of $\Delta\mu$ values assumed to be entirely describing the projected velocity of the photocenter around the barycentre 
at the \hip\ and \gdrtwo\ epochs. 

To include a time series of absolute astrometry for $\beta$ Cyg Aa in our analysis, we take the cross-calibrated \hip /\gdrtwo\ proper motion values 
and the scaled \hip -\gaia\ positional difference from the \citet{Brandt2018,Brandt2019b} catalogue of astrometric accelerations\footnote{\citet{Brandt2018,Brandt2019b} employs a linear 
combination of the original \citep{Perryman1997} and new \citep{NewHipp2007} reductions of the \hip\ data to estimate star positions, 
as a way to mitigate the systematics associated with each of the catalogues considered individually. We actually adopt the version of the catalogue 
published in \citet{Brandt2019b}, which corrects an error in the calculation of the perspective acceleration in R.A. and effectively supersedes the original 
catalogue presented in \citet{Brandt2018}.}. However, \citet{Brandt2018} warns against blind 
use of the $\Delta\mu$ technique in cases of accelerating stars that are binaries with modest brightness ratios and/or stars with particularly large errors in the 
astrometry. This is indeed the case for $\beta$ Cyg Aa: due to its brightness, it is highly saturated in \gaia\ data, and $\beta$ Cyg Ac is only 2.2 mag fainter 
at $V$ band (see Table \ref{Tab:PhysPar}). Furthermore, it is questionable whether the scaled \hip -\gaia\ positional difference, spanning only $\sim20\%$ of the 
orbital phase, can really be considered as a close representation of the barycentre tangential velocity. In our analysis, we then decide to utilize the three proper
motions separately\footnote{Following \citet{Lindegren2020,Lindegren2020b}, when calculating the \hip\ - \gaia\ position differences divided by the epoch difference 
we subtract the (small) cross-calibration corrections applied by \citet{Brandt2018,Brandt2019b} in order to place them on the ICRS. }. 
We report in Table \ref{tab:astrom_tbl} the positions and proper motions of both \bca\ and B from \hip , 
\gaia\ and Tycho-2, as well as \bca's cross-calibrated \hip\ and \gaia\ proper motions and the corresponding scaled position difference as derived 
by \citet{Brandt2018,Brandt2019b}. These values will be utilized in our final orbit solution of the \bca\ system, and in the discussion of its relation to \bcb. 

\section{Physical parameters from spectroscopy and evolution models}
\label{sec:spectro}

We present in Figures~\ref{fig_spec} and \ref{fig_spec2} the high resolution 
($R\approx 20,000$) spectrum obtained with the TIGRE telescope. The spectrum
contains contributions from both stars \bca a and \bca c. Blueward of 4000 Angstroms the Balmer series of the blue main-sequence companion \bca c can be clearly identified (Fig.~\ref{fig_spec}). The rest of the spectrum is dominated by the various lines of the giant star \bca a, the \ion{Ca}{ii} triplet being clearly seen redward of 8475 Angstroms (last row of Fig.~\ref{fig_spec2}).

\subsection{Albireo Aa, the red supergiant primary}

To improve our earlier spectroscopic 
determination of the physical parameters (see \citet{Jack2018}), we added 
all observed spectra to obtain 
a single spectrum with a very high S/N. Before this step, all spectra were 
corrected individually for their respective radial velocity against the 
laboratory wavelength scale. In this improved analysis we now also used 
some unblended lines recorded in the blue channel of HEROS.

To avoid confusion with the spectral contributions from the very near 
component Ac, which falls into the 3 arcseconds spanning aperture of 
the optical fibre feeding HEROS, and from line blanketing too strong 
to define a continuum, we restricted this work to lines redwards of 
4900~\AA. In the red channel of the HEROS spectrograph we also excluded 
all regions which are affected by telluric line absorption.

The parameter-fitting procedure was carried out with the spectral analysis 
tool-kit iSpec \citep{Blanco2014} in its most recent version (v2019.03.02), described by 
\citet{Blanco2019}. In that work, the creator of iSpec recommends the 
use of a list of specific lines, which are matching the solar spectrum very 
well for the known physical parameters of the Sun. However, using these 
lines in our work, we found that the suggested best-fit parameters depend 
on the choice of their initial values (which they should not), especially
when several parameters are optimized in the same run. Total 
error sums $\chi^2$ of such competing parameter-fits differ very little. 
We suspect that this is the effect of working here in a regime of lower 
temperature and gravity, as compared to the Sun.
Thus, we here did not use that suggested line list, but instead we started
with the GES line list \citep{Gilmore2012, Randich2013}, 
and in the wavelength range from 4900 to 8800~\AA\ we 
then selected those spectral segments without telluric contamination.

As suggested by \citet{Blanco2019}, we used the iSpec parameter-fitting 
option, which is based on the SPECTRUM code \citep{Gray1994}, and 
as solar abundances we chose the ones 
of \citet{Grevesse2007}. The atmospheric model spectra used by us in iSpec
for the best-fit comparison are of the MARCS code \citep{Gustafsson2008}.
To obtain a reasonable value for the rotation velocity,
we fixed the micro and macro turbulence velocities by the empirical relation
offered now by iSpec in its newest version.

%
\begin{table}
\centering
\caption{Stellar parameters of \bca a as determined by spectroscopic analysis}
\label{tab:Aa_parameters}
\begin{tabular}{lc}
\hline
Parameter &  Value\\
\hline
$T_\mathrm{eff}$ & $4382.7\pm2.1$~K  \\
$\log{g}$ & $0.93\pm0.01$ \\
$[M/H]$ & $0.02$  \\ 
$[\alpha/Fe]$ & $0.08$ \\
$v_\mathrm{mic}$ & 1.57~km\;s$^{-1}$ \\
$v_\mathrm{mac}$ & 5.22~km\;s$^{-1}$  \\
$v\,\sin{i}$ & $8.34\pm0.4$~km\;s$^{-1}$ \\
\hline
\end{tabular}
\end{table}

The results of this spectroscopic stellar parameter determination for \bca a 
by iSpec are listed in Table~\ref{tab:Aa_parameters}. Most importantly, 
the effective temperature (essentially unchanged from \citet{Jack2018}) 
of $T_\mathrm{eff}= 4382.7\pm2.1$~K will be used for the evolution modelling
to obtain mass and age of component Aa in the following section. We note that the quoted uncertainty of $T_\mathrm{eff}$ is the internal error (of the fitting process), and that the total uncertainty is more like 60 K (see \citet{Schroeder2020p}). The conclusion of \citet{Jack2018}, that \bca a is a normal giant star, 
below is confirmed and substantiated.
%
%
\begin{figure}
\includegraphics[width=\columnwidth]{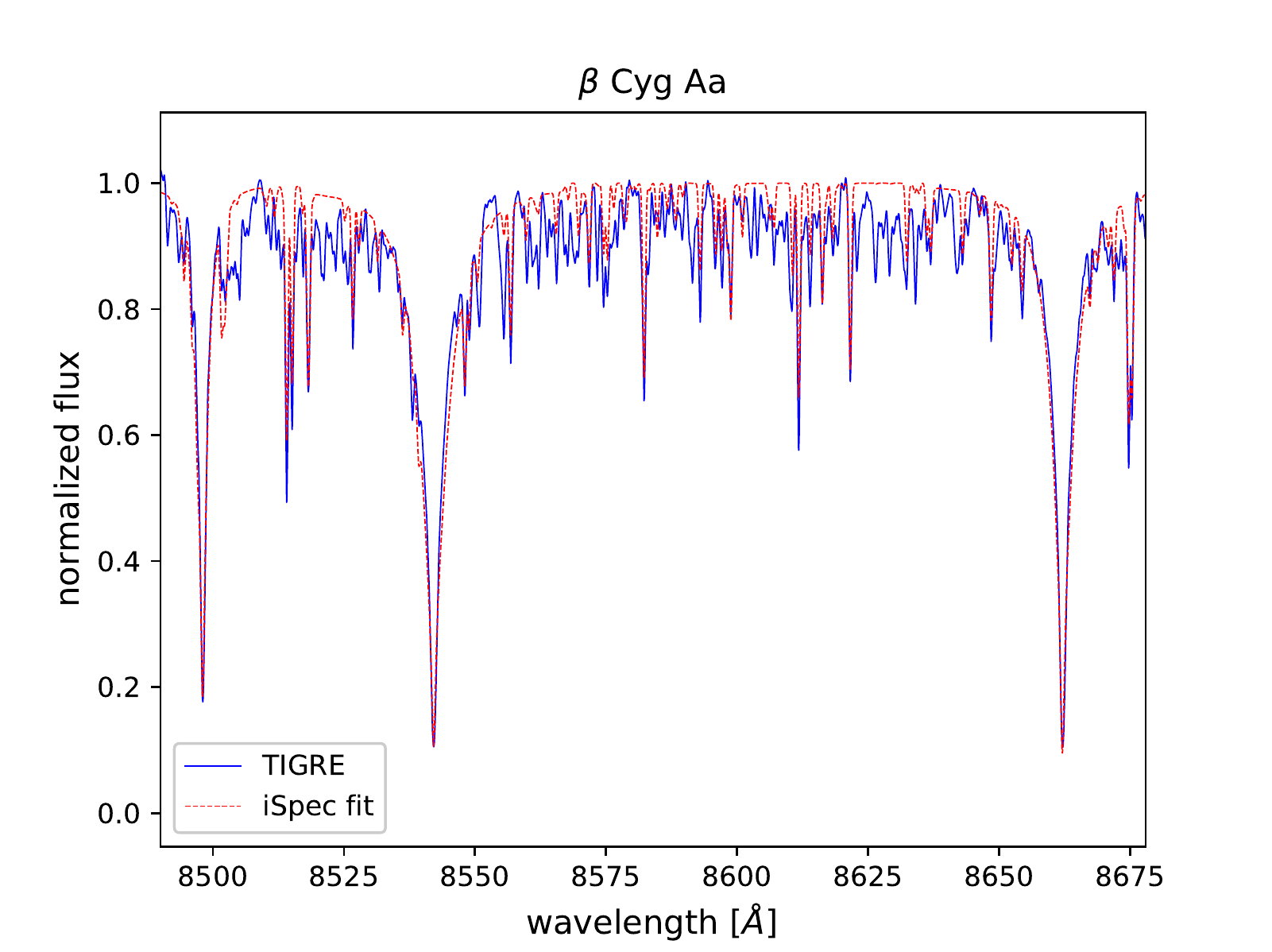}
\caption{The observed TIGRE spectrum with the iSpec fit in the region of the Ca triplet.}
\label{fig_specfit}
\end{figure}
%

There are some small differences to our previous results (\citet{Jack2018}). 
For example, the spectroscopic surface gravity here is slightly higher. This 
actually improves consistency with the gravity ($\log{g}=1.57$), which we 
compute from the mass of $5.2 M_{\odot}$ of the best-fit evolutionary 
model for Aa (see next section), and the radius of $62R_{\odot}$ 
resulting from the luminosity and effective temperature. In a work on the
Hyades K giants \citep{Schroeder2020}, we have already noted this same 
tendency. i.e. that iSpec best-fit gravities turn out about 0.5 dex lower 
than the computational value. However, cross-talk
with effective temperature values is still small on this scale: Setting
the gravity up to such a larger value, the value suggested by iSpec
for $T_{\rm eff}$ changes by less than 2\%.

As a new spectroscopic parameter, the $\alpha$ enhancement of Albireo Aa 
was obtained by us here as $+0.08$. And with the values for the micro and 
macro-turbulence taken from empirical relations, we determined a notable 
rotational velocity for the giant of $v\sin{i}=8.34\pm0.4$~km\;s$^{-1}$.  
It is worth mentioning that
the errors given in Table~\ref{tab:Aa_parameters} are the errors
of the fitting procedure and do not include any systematic errors, which 
we attempt to estimate and add in the following.

\subsubsection{Photometric properties}

In Table~\ref{Tab:PhysPar} we list the photometric properties of the primary
star Aa which result from subtraction of the small contributions of Ac
(see below) to the B and V magnitudes given by SIMBAD for the composite
light of Albireo A. They are followed by the respective luminosity.
Here we are using as the distance of \bca\ that of \bcb\ 
(120pc), for the reasons discussed in Section \ref{sec:data}.


As bolometric correction for the supergiant
primary Aa we use $\mathrm{BC}=-0.80$, as taken from \citet[see Fig 4 therein]{Flower1996}, 
which corresponds to the colour ($B-V=1.25$) and the effective temperature of Aa derived above. 
Apart from the debated parallax and true distance to \bca\, the
bolometric correction presents the other larger uncertainty in the
luminosity of Aa, since BC is changing steeply around the effective temperature of this
red supergiant.

\begin{table}
\caption{Physical parameters of the Albireo components}
\begin{center}
\begin{tabular}{r |r | r |r  } 
 \hline
    component & Aa & Ac & B \\ 
 \hline
            $V$          &  $3.21\pm0.04$ &  $5.85\:$  &  $5.11\pm0.02$ \\
	        $B$          &  $4.45\pm0.04$ &            &  $5.01\pm0.02$ \\
          $B-V$          &  $1.25\pm0.05$ &            & $-0.10\pm0.03$ \\
 \hline
           BC            & $-0.80\pm0.10$ & $-0.40\pm0.10$ & $-0.85\pm0.10$ \\
           $M_{\rm Bol}$ & $-3.01\pm0.13$ & $+0.05:      $ & $-1.16\pm0.12$ \\
 \hline
           $\log{L/L_{\odot}}$ & $3.10\pm0.05$ & $1.9\pm0.1:$ & $2.36\pm0.05$ \\
           $\log{T/K}$         & $3.64\pm0.02$ & $4.0\pm0.05$ & $4.121\pm0.02$ \\
           $M/M_{\odot}$       & $5.2\pm0.1$   & $2.7\pm0.1$ & $3.7\pm0.05$ \\ 
\hline
\end{tabular}
\end{center}
\label{Tab:PhysPar}
\end{table}

\subsection{The very close secondary Ac}
Like Albireo B, the close companion Ac is a hot main-sequence star.
Since its separation from the bright primary Aa is less than
an arcsecond in the sky, Ac is in the seeing-related glare of Aa.
Consequently, the assessment of the physical parameters of Ac
is severely affected. This being a physical binary as described
above, and so the distance of Ac without doubt should be same as
the one adopted above for Aa (120 pc). The CCDM entry of Ac
suggests V=5.5 but without any reference. A work dedicated to close
binary components by \citet{tenBrummelaar2000}, using speckle differential
photometry on the venerable 100 inch Mt. Wilson reflector, measured 
a visual magnitude difference between Aa and Ac of 2.64 mag, bringing 
Ac to a more credible V=5.85 mag. In addition, the spectral type of B9.5V 
would suggest a B-V value of close to 0.0, and an effective temperature 
of close to 10,000 K, but that we must regard as uncertain.

IUE took a low-resolution SWP spectrum of Albireo. Using the
larger (10 arcseconds wide) slit, in which the small displacement
of Ac against Aa is irrelevant. The giant component Aa already
being very weak in the far UV, this spectrum seemed promising for
getting additional information on effective Temperature and luminosity of
Ac. However, comparing its SED (spectral energy distribution) with PHOENIX
models, the best-matching $T_{\rm eff}$ remains ambigious. While the very
notable flux beyond Ly-alpha, i.e. shortward of 121nm, suggests Ac to be
hotter than 11,000 K,
the long wavelength end of the SED (170-220 nm) has a slope, which is better
consistent with spectra of 10,200 K models. Within this uncertainty in
$T_{\rm eff}$, which quadrupels into the respective luminosity estimate,
we can only say that the IUE SWP fluxes are consistent with a 5.85 magnitude
A0-B9 main sequence star at the distance of Albireo within a factor of 2.
Consistent with the assumption that Ac is still near the Zero-Age Mainsequence,
the synthetic spectra used here are of non-LTE PHOENIX models
with $\log{g}=5.0$ and solar metallicity, obtained from
the PHOENIX model library, made publically available
by the University of G\"ottingen and described by \citet{Husser2013}.
To match the IUE SWP low dispersion, we reduced the resolution of
the synthetic spectra with iSpec.


\subsection{Albireo B, the far hot companion}
Albireo B is separated from Aa by 34 arcseconds, and it is a bright
($m_V=5.11$~mag) star in its own right. Not surprisingly then, it has been studied
well in the past. We adopt $T_{\rm eff}$=13,200 K according to
\citet{Levenhagen2004} and consistent with the spectral type of B8V.
With a BC of $-0.85$ (according to \citet{Flower1996}, see Fig. 4 therein),
and a distance of 120 pc, we then obtain a luminosity of 229 $L_{\odot}$
($\log{L/L_{\odot}}=2.36$, see Table~\ref{Tab:PhysPar}).

Again, as with Aa, the bolometric correction here seems to be the other main source of uncertainty, since
Albireo B, too, lies in a temperature range where BC changes steeply.

\subsection{Masses by evolution modelling - same age?}
\label{sec:evolution}


Matching the physical quantities of each of the Albireo component stars
with an evolution track in the HRD (i.e. $\log{T_{\rm eff}}$ and
$\log{L/L_{\odot}}$) reveals their masses and  ages.
For this work, we are using the well-tested evolution models
of the Eggleton code (see \citet{Schroder1997} and \citet{Pols1997}),
where the amount of "overshooting" (in better words: extra mixing)  and
the mixing length itself was carefully calibrated against eclipsing
binaries and stellar cluster isochrones in the mass-range relevant
for this study.

If Albireo was once formed as a hierarchical triple, or may even still be
one (see below), then all three stars should be matched with models of
the same age. A star on the main sequence, like the companions
Ac and B, changes its HRD position only slowly, and so its match is less
critical. On the contrary, giant star evolution is fast and therefore a
sensitive indicator of age.

\begin{figure}
\includegraphics[width=1.0\linewidth]{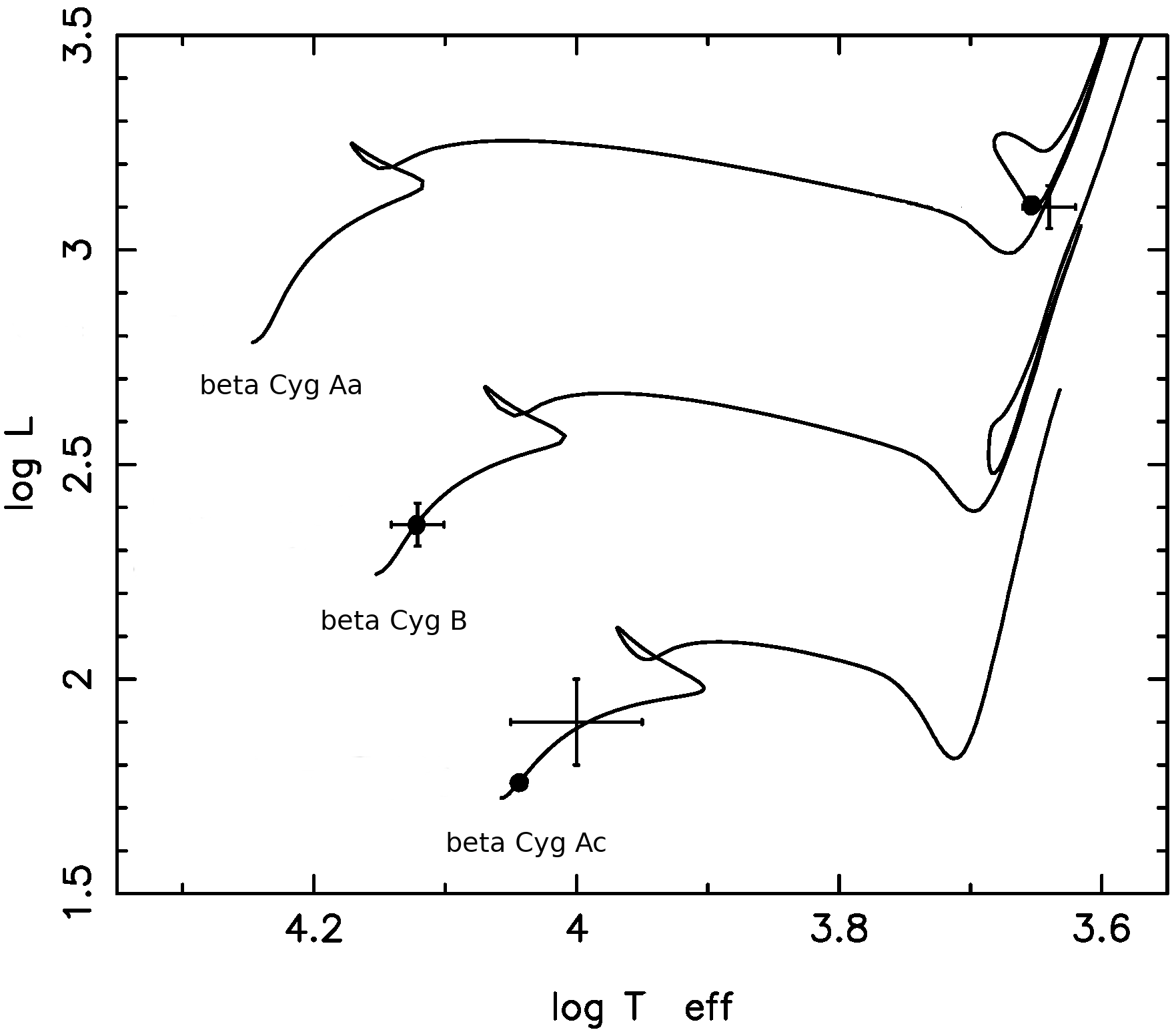}
\caption{Evolutionary tracks with the suggested masses of the three Albireo components,
of 5.2 $M_{\odot}$ for the primary (Aa), 3.7 $M_{\odot}$ for the wide
companion Albireo B, and 2.7 $M_{\odot}$ for the close secondary (Ac). 
(Luminosity is with respect to solar, and the effective temperature is in Kelvin.) The
age of 99 Myrs is best defined by the primary, which is starting central
helium burning, and Albireo B matches that age mark very well, Ac still
within its larger error bars. The physical parameters and their estimated
uncertainties are as listed in Table~\ref{Tab:PhysPar}.}
\label{fig_evoltracks}
\end{figure}

For the primary, Albireo Aa, we find a very good match with an evolution track
of a mass of 5.2$M_{\odot}$, as just starting its central  He-burning
(see Fig.~\ref{fig_evoltracks}). Since the abundances of Albireo are not
much different from solar, we computed that track (and all others shown
here) for a metallicity of Z=0.02.

The early central helium burning is about the slowest phase in the
evolution of a red giant with a few solar masses. By contrast, a star with the mass of Albireo Aa passes its shell-burning
phases quite fast. Therefore, the onset of central helium-burning
has the largest probability, compared with alternative evolutionary tracks,
which would --- within the uncertainties --- match the present HRD location.
At the same time this solution points to a well-defined age for the Albireo primary,
of 99 Myrs.

Within their uncertainties discussed above (and see Table~\ref{Tab:PhysPar}),
the two companions are matched well with evolution tracks of 3.7$M_{\odot}$ 
for Albireo B and 2.7$M_{\odot}$ for the close secondary Ac, see
Fig.~\ref{fig_evoltracks}, while coinciding with the age of Aa. 
We would like to emphasize the perfect age match of the distant companion
Albireo B with the primary Aa. This may not prove
a common origin as a hierarchical triple, but it supports this
idea to be, at the least, very likely. 

In the case of Ac, this coincidence looks near marginal, though, as if Ac was
already more evolved, that is: having a larger content of Helium that the age
of Aa and an undisturbed evolution as a single star would suggest. 
But by means of the orbit there can be no doubt, that Albireo 
Aa/Ac is a bound system, which was formed together. To resolve this issue, as well
as the larger system mass and mass ratio demanded by astrometry, when compared to 
the here given astrophysical account of the visible mass, see further 
suggestions on the nature of Ac in our discussion below.

\section{Orbit solutions}
\label{sec:orbit}
\subsection{Preliminary RV orbital solutions}
\label{sec:RVonly_orbit}

As a first step in the derivation of an updated orbital solution for the \bca\ system, we analysed all the RV data available (described in Section~\ref{sec:RVdata}) with tools commonly used in the analysis of RV extrasolar planets \citep[e.g.][]{pinamontietal2018}. This is a particularly interesting case for such an analysis, since Albireo has one of the longest baselines of RV data ever collected.

The model adopted to describe the radial velocity time series is the following:
\begin{equation}
\label{eq:rv_model}
RV_\text{mod} = \gamma_\text{syst} + K \cos(\nu(t, e, T_0, \textit{P}) + \omega) +  e\cos(\omega),
\end{equation}
where $\gamma_\text{syst}$ is the systemic radial velocity of the system, $K$ is the semi-amplitude of the Keplerian signal, and the true anomaly $\nu$ is a function of time \textit{t}, time of the inferior conjunction $T_0$, orbital period $P$, eccentricity $e$ and argument of periastron $\omega$.
We choose not to consider different RV offset values for the different instruments, since most time series have very short timespans compared to the expected orbital period of the binary, and thus the additional degree of freedom given by the offset would almost void the contribution of the short-baseline datasets to the fit.\footnote{We explored the addition of instrumental offsets to the RV-only model, but the test fit (not shown), resulted in uncertainties almost as large as the priors.} Moreover, instead of fitting separately the eccentricity $e$ and argument of periastron $\omega$, we define the auxiliary parameters $e \cdot \cos(\omega)$ and $e \cdot \sin(\omega)$ in order to reduce the covariance between the fitted parameters.

The model described in Eq. \ref{eq:rv_model} is fitted via MCMC analysis, performed with the publicly available \texttt{emcee} algorithm \citep{foreman13}. We used 100 random walkers to sample the parameter space. The posterior distributions were derived after applying a burn-in phase of 3000 steps, as explained in \citet[and references therein]{eastman13}.
To evaluate the convergence of the different MCMC analyses, we calculated the integrated correlation time for each of the parameters, and stopped the code after a number of steps equal to 200 times the largest autocorrelation times of all the parameters \citep{foreman13}.

\begin{figure}
   \centering
   \includegraphics[width=.5\textwidth]{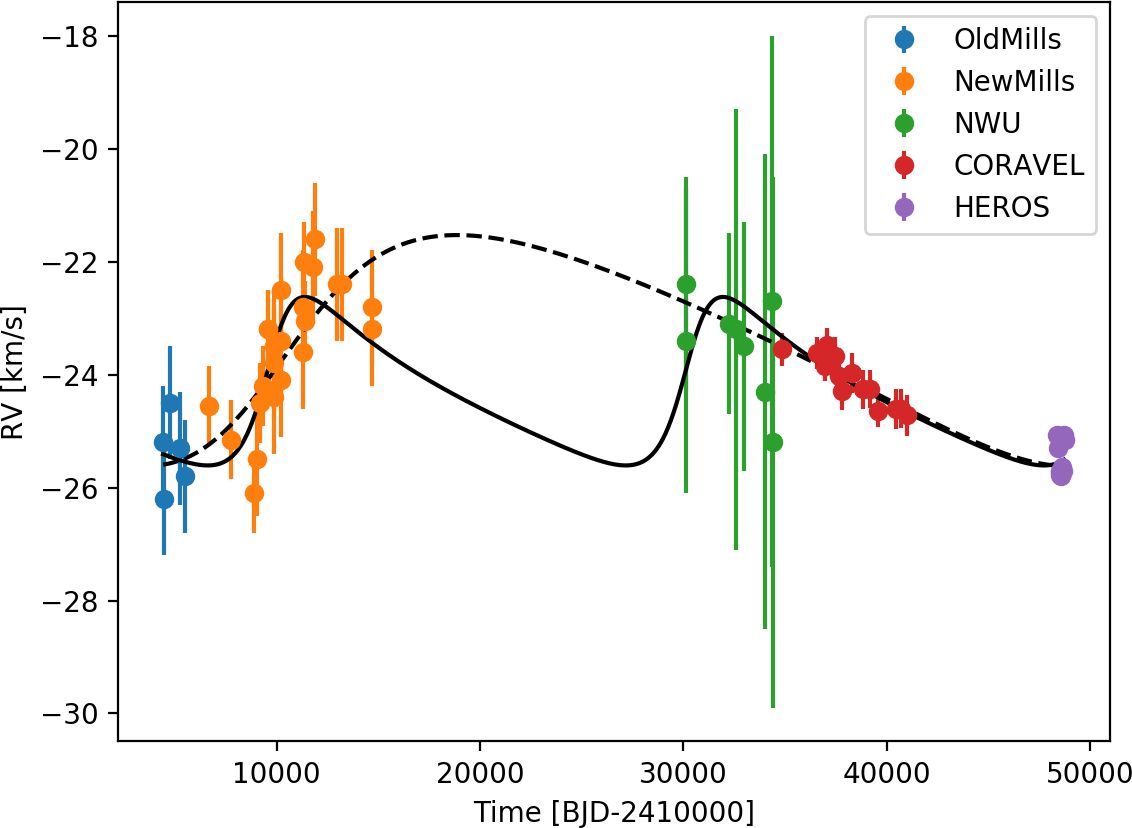} 
      \caption{Short- and long-period orbital solutions for the Aa-Ac system compared with the five radial velocity time series considered in the analysis. Color of the data points notes the telescope/instrument used, as noted in Table \ref{tab:RVs_tbl}, where LOM = Lick/Original Mills, LNM = Lick/New Mills, NWU= NWU LARC, HPO = HPO/CORAVEL, KIT = Kitt Peak, and HEROS = TIGRE/HEROS. }
         \label{fig:rv_model}
\end{figure}

We tested different initial configurations for the \texttt{emcee} code as well as different priors for the orbital parameters, in order to test the different orbital solutions from the literature \citep{Hartkopf1999,Scardia2007}. Long-period and short-period solutions were treated separately, since the \texttt{emcee} code is susceptible to multi-modal probability distributions \citep{foreman13}.
We found the best-fit solutions to be the two obtained with the priors listed in Table \ref{tab:rv_model}, and shown in Fig. \ref{fig:rv_model}, respectively one for a short-period orbit, and one for a long-period orbit. The best fit results are described by the median of the distribution and the asymmetric error bars obtained from the 16th-84th percentiles.

\begin{table}
   \caption[]{Priors and best-fit results for the \texttt{emcee} analysis of the combined RV time series of \bca a, for a Keplerian model of the Aa/Ac system.}
          \label{tab:rv_model}
          \centering
         \tiny
    \begin{tabular}{l l l l l}
             \hline
             \noalign{\smallskip}
             & \multicolumn{2}{c}{Short period} & \multicolumn{2}{c}{Long period}  \\
             parameter     &  Prior & Best-fit value &  Prior & Best-fit value  \\
             \hline
             \noalign{\smallskip}
             $\gamma_\text{syst}$ [km s$^{-1}$]  & $\mathcal{U}$(-30,-20) & -24.33$^{+0.10}_{-0.13}$  & $\mathcal{U}$(-30,-20) & -23.59$^{+0.21}_{-0.20}$ \\
             \noalign{\smallskip}
             $K$ [km s$^{-1}$] & $\mathcal{U}$(0,4) & 1.40$^{+0.20}_{-0.24}$ & $\mathcal{U}$(0,4) & 1.84$^{+0.30}_{-0.23}$ \\
             \noalign{\smallskip}
             $P$ [years] & $\mathcal{U}$(40,100) & 55.7$^{+2.0}_{-1.1}$ & $\mathcal{U}$(105,150) & 120.5$^{+15.3}_{-7.4}$ \\
             \noalign{\smallskip}
             $T_\text{0}$ [BJD-2410000] & $\mathcal{U}$(0,50000) & 18000$^{+1400}_{-1500}$ & $\mathcal{U}$(0,50000) & 36200$^{+2200}_{-1900}$ \\
             \noalign{\smallskip}
             $\sqrt{e} \cdot \cos\omega$ & $\mathcal{U}(-1.0,1.0)$ & 0.11$^{+0.16}_{-0.22}$ & $\mathcal{U}(-1.0,1.0)$ & -0.20$^{+0.19}_{-0.14}$ \\
             \noalign{\smallskip}
             $\sqrt{e} \cdot \sin\omega$ & $\mathcal{U}(-1.0,1.0)$ & -0.69$^{+0.13}_{-0.09}$ & $\mathcal{U}(-1.0,1.0)$ & -0.58$^{+0.15}_{-0.14}$ \\
             \noalign{\smallskip}
             \textit{Derived quantities} \\
             \noalign{\smallskip}
             $e$ &  & 0.52$^{+0.13}_{-0.14}$ &  & 0.40$^{+0.20}_{-0.15}$ \\
             \noalign{\smallskip}
             $\omega$ [rad] &  & -1.40$^{+0.26}_{-0.32}$ &  & -1.89$^{+0.30}_{-0.24}$ \\
             \noalign{\smallskip}
             \hline
      \end{tabular}
\end{table}

As shown in Fig. \ref{fig:rv_model}, the presented solutions follow closely the RV data of the binary system. Moreover, we compared the rms of the RV residuals, after the subtraction of $\gamma_\text{syst}$ and Keplerian signal, with the mean uncertainty $\langle \sigma_i \rangle$ of each dataset $i$: we obtained $\text{rms} / \langle \sigma_i \rangle < 0.9$ for all datasets. This is further evidence of the goodness of the two fits, since no excess noise is present in the RV residuals. 
However, it is impossible to distinguish between long- and short-period solutions: adopting the Bayesian information criterion \citep[BIC,][]{schwarz1978} we obtained $\Delta$BIC $= 0.17$ between the two models, which corresponds to no statistical evidence in favour of one or the other. This is an indication of the fact that the RV data alone, due to the short baseline compared to the orbital period, and the sparse sampling, cannot give a precise measurement of the orbital parameters of the \bca\ system on their own.

\subsection{Orbital solution from RVs and relative astrometry}
\label{sec:RVspeckle_orbit}

A Bayesian analysis of the combined RV and speckle observations was performed using a differential evolution Markov chain Monte Carlo (DE-MCMC) method \citep{Ter2006, eastman13}. 
We took advantage of the six ($A$, $B$, $C$, $F$, $G$, $H$) Thiele-Innes constants representation (e.g. \citet{Binnendijk1960, Wright2009}) to partially linearize the problem 
in both astrometry and RV. Within this dimensionality reduction scheme, 
only three non-linear orbital parameters must be effectively explored using the DE-MCMC algorithm (e.g., \citet{Casertano2008, Wright2009, Mendez2017}), 
namely $P$, $T_0$, and $e$. At each step of the DE-MCMC analysis, the resulting linear system of equations is solved in terms of the Thiele-Innes constants using 
simple matrix algebra, singular value decomposition and back-substitution being the method of choice. Standard formulae are then applied 
(e.g., \citet{Casertano2008, Wright2009, Mendez2017}) to convert from the Thiele-Innes constants back to the remaining Campbell elements (semi-major axis $a$, inclination angle $i$, 
argument of periastron $\omega$, and longitude of the ascending node $\Omega$), $K$, and $\gamma_\mathrm{syst}$. The speckle imaging astrometry and RV time series 
were modelled using the following likelihood function: 
\begin{eqnarray}
\ln \mathcal{L} &=& -\frac{1}{2} \left( \sum_{i=1}^{N_{RV}}\left( \frac{RV_i^{(obs)} - RV_i^{(model)}}{\sigma_{RV,i}}\right)^2 \right. \nonumber \\
                & & \left. + \sum_{j=1}^{N_{astr}}\left(\frac{X_j^{(obs)} - X_j^{(model)}}{\sigma_{X,j}}\right)^2 \right. \nonumber \\
                & & \left. +  \sum_{j=1}^{N_{astr}}\left(\frac{Y_j^{(obs)} - Y_j^{(model)}}{\sigma_{Y,j}}\right)^2 \right)
\end{eqnarray}
where $X = \rho \cos(PA)$ and $Y = \rho \sin(PA)$ are the relative positions from the speckle data, $N_{RV}$ and $N_{astr}$ are the number of RV and astrometric measurements, respectively, and $\sigma$ the corresponding measurement errors. At this stage, we did not consider 
offsets between datasets obtained by different instruments. The DE-MCMC analysis was carried out with a number of chains equal to twice the number of free parameters, and it was stopped after it reached convergence and good mixing of the chains based on the Gelman-Rubin 
statistics (e.g., \citet{Ford2006}). After removing 20\% of burn-in steps, the medians of the posterior distributions and their $\pm34.13\%$ intervals were evaluated and 
were taken as the final parameters and associated $1\sigma$ uncertainties. The final results are reported in Table \ref{tab:combined_model}. 
Overall, the combination of RV and speckle imaging astrometry allows to constrain rather robustly the orbital configuration of the Albireo Aa,c components. 
As a cross-check, the $\omega$ value derived from the Thiele-Innes representation for the RV model (describing the primary orbital motion) differs exactly by 
180 deg from the one obtained from the astrometry model (that describes the secondary orbit). 
We also explored solutions in the neighbourhood of the short ($\sim55$ yr) and long ($\sim210$ yr) period orbits derived with RV-only data in Sect. \ref{sec:RVonly_orbit} 
and by \citet{Scardia2007}, respectively, but these attempts returned lower likelihoods, and thus disfavoured solutions with respect to the one reported in Table \ref{tab:combined_model}.

Finally, assuming a distance of 120 pc (i.e. a parallax of 0.00833 arcsec), and given the values of $a$ (in arcsec), $P$ (in yr), $e$, $i$, and $K$ (in km s$^{-1}$) 
from Table \ref{tab:combined_model}, we can derive the mass ratio $q=M_\mathrm{Ac}/M_\mathrm{Aa}$ from e.g. Eq. 12 in \citet{Pourbaix2000}):
\begin{equation}
\frac{1}{1+1/q}= 0.0335729138\frac{P K\sqrt{1-e^2}\varpi}{a\sin i}.
\end{equation}
We obtain $q=0.58_{-0.10}^{+0.13}$. At the same distance, the total system mass can be inferred to be $7.58_{-0.46}^{+0.48}$ solar masses. 

\begin{table}
   \caption[]{Priors and best-fit results for the DE-MCMC analysis of the combined RV + speckle imaging time series.}
          \label{tab:combined_model}
          \centering
         \small
    \begin{tabular}{l l l}
             \hline
             \noalign{\smallskip}
             Parameter     &  Prior & Best-fit value \\
             \hline
             $P$ [yr] & $\mathcal{U}$(100,150) & 123.59$^{+2.86}_{-2.45}$  \\
             \noalign{\smallskip}
             $T_\text{0}$ [yr] & $\mathcal{U}$(2000.0,2050.0) & 2025.62$^{+0.86}_{-1.01}$  \\
             \noalign{\smallskip}
             $e$ & $\mathcal{U}$(0.0,0.99) & 0.204$^{+0.013}_{-0.014}$  \\
             \noalign{\smallskip}
             \textit{Derived quantities:} \\
             \noalign{\smallskip}
             $a$ [arcsec] &  & 0.406$^{+0.006}_{-0.005}$  \\
             \noalign{\smallskip}
             $i$ [deg] &  & 155.53$^{+2.68}_{-2.33}$  \\
             \noalign{\smallskip}
             $\Omega$ [deg] &  & 86.89$^{+4.64}_{-3.64}$  \\
             \noalign{\smallskip}
             $\omega$ [deg] &  & 54.04$^{+1.70}_{-1.88}$ \\
             \noalign{\smallskip}
             $\gamma_\text{syst}$ [km s$^{-1}$]  & & -23.56$^{+0.08}_{-0.09}$  \\
             \noalign{\smallskip}
             $K$ [km s$^{-1}$] &  & 1.81$^{+0.07}_{-0.08}$  \\
             \noalign{\smallskip}
             \hline
      \end{tabular}
\end{table}


\subsection{Final orbit solution}
\label{sec:finalOrbit}

As discussed in Section \ref{sec:absolute_ast}, the inclusion of the available information on the orbital motion of $\beta$ Cyg Aa coming from absolute astrometry 
allows one to obtain an estimate of the mass ratio in addition to the complete set of orbital elements. For the reasons outlined in Section \ref{sec:absolute_ast}, 
we decide to proceed using the three individual values of absolute proper motion rather than two proper motion differences. As a consequence, we need to solve 
not only for the mass ratio ($q$) but also for the two components of the proper motion of the barycentre of the system ($\mu_\mathrm{RA}^{b}$, $\mu_\mathrm{DEC}^{b}$). 
The DE-MCMC orbital fit analysis is then carried out on the time-series of absolute proper motions so defined, the RV and speckle imaging datasets, 
adding $q$, $\mu_\mathrm{RA}^{b}$, and $\mu_\mathrm{DEC}^{b}$ as new model parameters effectively explored using the DE-MCMC algorithm
\footnote{For the purpose of this study, in the modelling of the \hip -\gaia\ absolute astrometry of $\beta$ Cyg Aa we ignore any wavelength-dependent effects 
on the photocenter due to the presence of $\beta$ Cyg Ac}. At this stage, we allow for a free offset parameter for each of the five independent RV datasets. 
The final likelihood function used in the DE-MCMC analysis is: 
\begin{eqnarray}
\ln \mathcal{L} &=& -\frac{1}{2} \left( \sum_{l=1}^{N_{inst}}\sum_{i=1}^{N_{RV}}\left( \frac{RV_i^{(obs)} + RV_l^{(off)} - RV_i^{(model)}}{\sigma_{RV,i}}\right)^2 \right. \nonumber \\
                & & \left. + \sum_{j=1}^{N_{astr}}\left(\frac{X_j^{(obs)} - X_j^{(model)}}{\sigma_{X,j}}\right)^2 \right. \nonumber \\
                & & \left. + \sum_{j=1}^{N_{astr}}\left(\frac{Y_j^{(obs)} - Y_j^{(model)}}{\sigma_{Y,j}}\right)^2 \right. \nonumber \\
                & & \left. + \sum_{k=1}^{3}\left(\frac{\mu_{\mathrm{RA},k}^{(obs)}- \mu_{\mathrm{RA},k}^{(model)}}{\sigma_{\mu_{\mathrm{RA},k}} }\right)^2 \right. \nonumber \\
                & & \left. + \sum_{k=1}^{3}\left(\frac{\mu_{\mathrm{DEC},k}^{(obs)}- \mu_{\mathrm{DEC},k}^{(model)}}{\sigma_{\mu_{\mathrm{DEC},k}}}\right)^2 \right )
\end{eqnarray}

\begin{figure*}
   \centering
   \includegraphics[width=.75\textwidth]{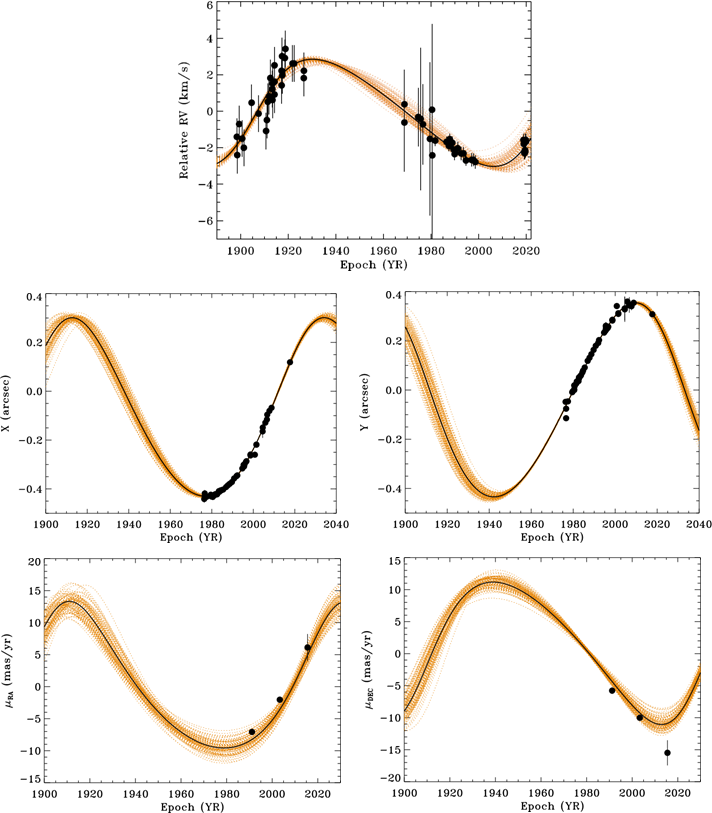} \\
      \caption{The best-fit Keplerian model (solid black line) superposed to the RV time series (top panel), to the speckle imaging data (middle panels), 
      and to the absolute proper motions from \hip\ and \gaia\ (lower panel). The dashed orange lines represent a random selection of orbit solutions 
      drawn from our DE-MCMC posteriors.}
         \label{fig:bestfit_combined_full_model}
\end{figure*}

The DE-MCMC analysis was carried out in the same way as described in the previous section. The final results are reported in Table \ref{tab:combined_full_model} 
and Figures \ref{fig:joint_posteriors_full_model}, \ref{fig:jump_parameters_posteriors_full_model} 
and \ref{fig:derived_parameters_posteriors_full_model}, with Figure \ref{fig:bestfit_combined_full_model} showing the best-fit solutions for the RV time-series, 
speckle data (in cartesian coordinates), and absolute astrometry. 

The combination of radial velocity data, relative and absolute astrometry is the most complete dataset available to us and the orbital solution presented 
in Table \ref{tab:combined_full_model} is the one we consider as final. 
The orbital elements of the $\beta$ Cyg A system (both fitted and derived) are all determined within 1$\sigma$ of the values obtained by fitting an orbit only 
to RVs and speckle imaging data. The orbital configuration is therefore robustly confirmed based on the combined modelling of the three datasets. 
Also in this case, attempts at fitting shorter- or longer-period orbits as done in Sect. \ref{sec:RVspeckle_orbit} confirmed that the solution reported 
in Table \ref{tab:combined_full_model} is the one with the higher likelihood.

To further quantify the quality of the global fit, we use the same statistics utilized in Sect. \ref{sec:RVonly_orbit}, i.e. the ratio of the rms of the residuals of a given 
dataset to its mean uncertainty: A value close to unity is an indication that the residuals are fully compatible with the reported measurement uncertainties. 
For the RV datasets, this ratio is always $<0.85$. For the speckle imaging data, we have 0.86 and 1.44 along the X- and Y-axis, respectively. For the absolute 
astrometry we obtain 0.23 and 2.68 in RA and DEC, respectively. The only discrepant value is the proper motion in DEC at the mean \gdrtwo\ epoch, not unexpected 
given the large uncertainties in the astrometric solution for such a bright star. 

The values of systemic proper motion and mass ratio are highly correlated, as expected. The $\mu_\mathrm{DEC}^{b}$ value is compatible at the $1.6\sigma$ level 
with the equivalent long-term value for the star in the Tycho-2 catalogue \citep{Hog2000}, while the $\mu_\mathrm{RA}^{b}$ value obtained in our solution and the 
Tycho-2 one are discrepant at the $6.7\sigma$ level. In contrast to the estimate made in the previous subsection, the $q$ value we obtain in our global solution 
is distance independent. We find $q=1.25^{+0.19}_{-0.17}$. This value is surprising, as it points to a much larger mass for Albireo Ac than previously thought, 
or to its possible binarity. As a consequence, the $K$-value is also significantly larger than the one obtained in the solution presented in the previous section 
(Table \ref{tab:combined_model}), in part possible now that we have introduced zero-point offsets for each RV dataset, which are found to be within 
$1-2$ km s$^{-1}$ of the median RV value of each dataset. An additional effect is that the uncertainty on the systemic radial velocity $\gamma_\text{syst}$ 
(reported in Table \ref{tab:combined_full_model} as the weighted average of the individual RV zero-points) now realistically includes possible calibration errors. 

As shown in Figure \ref{fig:derived_parameters_posteriors_full_model}, the posterior distributions of parallax and total system mass allow us to infer directly from 
the data that $\varpi = 7.75_{-1.23}^{+1.25}$ mas (and thus $d=129.01_{-17.95}^{+24.39}$ pc) and $M_\mathrm{tot} =  9.47_{-2.23}^{+5.88}$ M$_\odot$ 
(with individual component masses $M_\mathrm{Aa} = 4.21_{-1.57}^{+2.87}$ M$_\odot$ and $M_\mathrm{Ac} = 5.23_{-1.69}^{+3.08}$ M$_\odot$).

\begin{table}
   \caption[]{Priors and best-fit results for the DE-MCMC analysis of the combined RV + speckle imaging + absolute astrometry time series. 
   The value of $\gamma_\text{syst}$ is taken as the median of the prior distribution of all the RV offsets.}
          \label{tab:combined_full_model}
          \centering
         \small
    \begin{tabular}{l l l}
             \hline
             \noalign{\smallskip}
             Jump parameter     &  Prior & Best-fit value \\
             \hline
             \noalign{\smallskip}
             $P$ [yr] & $\mathcal{U}$(100,150) & 121.65$^{+3.34}_{-2.90}$  \\
             \noalign{\smallskip}
             $T_\text{0}$ [yr] & $\mathcal{U}$(2000.0,2050.0) & 2026.36$^{+1.18}_{-1.04}$  \\
             \noalign{\smallskip}
             $e$ & $\mathcal{U}$(0.0,0.99) & 0.20$^{+0.01}_{-0.02}$  \\
             \noalign{\smallskip}
             $q$ & $\mathcal{U}$(0.0,2.0) & 1.25$^{+0.19}_{-0.17}$  \\
             \noalign{\smallskip}
             $\mu^{b}_\mathrm{RA}$ [mas yr$^{-1}$] & $\mathcal{U}$(-10.0,10.0) &  1.22$^{+0.22}_{-0.23}$  \\
             \noalign{\smallskip}
             $\mu^{b}_\mathrm{DEC}$ [mas yr$^{-1}$] & $\mathcal{U}$(-10.0,10.0) & $-$0.17$^{+0.64}_{-0.65}$  \\
             \noalign{\smallskip}
             $RV_1^{off}$  & $\mathcal{U}$(-50.0,0.0) & $-$23.79$^{+0.15}_{-0.10}$  \\
             \noalign{\smallskip}
             $RV_2^{off}$  & $\mathcal{U}$(-50.0,0.0) & $-$25.01$^{+0.26}_{-0.10}$  \\
             \noalign{\smallskip}
             $RV_3^{off}$  & $\mathcal{U}$(-50.0,0.0) & $-$22.78$^{+0.28}_{-0.52}$  \\
             \noalign{\smallskip}
             $RV_4^{off}$  & $\mathcal{U}$(-50.0,0.0) & $-$21.94$^{+0.14}_{-0.33}$  \\
             \noalign{\smallskip}
             $RV_5^{off}$  & $\mathcal{U}$(-50.0,0.0) & $-$23.49$^{+0.71}_{-0.59}$  \\
             \noalign{\smallskip}
             \textit{Derived quantities} \\
             \noalign{\smallskip}
             $a$ [arcsec] &  & 0.401$^{+0.007}_{-0.006}$  \\
             \noalign{\smallskip}
             $i$ [deg] &  & 156.15$^{+2.90}_{-2.63}$ \\
             \noalign{\smallskip}
             $\Omega$ [deg] &  & 84.43$^{+5.27}_{-4.50}$ \\
             \noalign{\smallskip}
             $\omega$ [deg] &  & 54.72$^{+1.88}_{-2.24}$ \\
             \noalign{\smallskip}
             $\gamma_\text{syst}$ [km s$^{-1}$]  & & -23.54$^{+1.43}_{-1.33}$ \\
             \noalign{\smallskip}
             $K$ [km s$^{-1}$] &  & 2.91$^{+0.09}_{-0.12}$ \\
             \noalign{\smallskip}
             $\varpi$ [mas] &  & 7.75$_{-1.23}^{+1.25}$  \\
             \noalign{\smallskip}
             $d$ [pc] &  & 129.01$_{-17.95}^{+24.39}$  \\
             \noalign{\smallskip}
             $M_\mathrm{tot}$ [M$_\odot$] &  & $9.47_{-3.24}^{+5.88}$ \\
             \noalign{\smallskip}
             $M_\mathrm{Aa}$ [M$_\odot$] &  & $4.21_{-1.57}^{+2.87}$ \\
             \noalign{\smallskip}
             $M_\mathrm{Ac}$ [M$_\odot$] &  & $5.23_{-1.69}^{+3.08}$ \\
             \noalign{\smallskip}
             \hline
      \end{tabular}
\end{table}

\section{Dynamical analysis of the Albireo system}
\label{sec:dynamical}

\subsection{Direct-measurement approach}

Can astrometric or radial-velocity measurements be used to directly confirm that the Albireo triple is gravitationally bound? 
Let us assume the best-case scenario that all three stars are at the same distance, and that: 
\begin {itemize} 
\item the distance of the system is 120\,pc (Section \ref{sec:absolute_ast}). 
\item the masses of the partners are as derived in Section\,\ref{sec:spectro} above, i.e.~that the total mass of the system is 11.6\,M$_\odot$,
\item the separation of the wide pair in the line-of-sight spatial coordinate is equal to the average separation in the two sky plane coordinates, i.e.~2927~$au$, (corresponding to 34.5\,arcsec/$\sqrt(2)$) at 120\,pc,
\item and that the present 3-dimensional separation following from the above assumptions (namely 5070\,$au$ or 0.025\,pc) is the semi-major axis of the orbit. 
\end{itemize}
Then Kepler's laws give the following measurable parameters (some of them depending on inclination, eccentricity and orbital phase):
\begin {itemize} 
\item an orbital period of around 106\,000 years
\item an orbital velocity in the order of 1.4\kms, of which on average a proportion of $1/\sqrt(3)$, i.e.~0.8\kms would point in the line-of-sight direction (at least in certain orbital phases) 
\item a relative orbital proper motion in the order of 2\,\masyr   
\item a relative angular acceleration in the order of 0.12\,$\mu$as yr$^{-2}$  
\item an orbital acceleration of 5\,mm s$^{-1}$ yr$^{-1}$.  
\end{itemize}

We note that the difference between the systemic proper motion of \bca\ and the proper motion of B is indeed of the order of 2\,\masyr. However, as we will see below, the observed difference in the radial velocities of A and B still make it unlikely that these two stars are bound. 

\subsection{Statistical approach}
\label{sec:statisticalApproach}


Another approach we can take to study the dynamical state of Albireo is a statistical one. As discussed earlier in this paper, we do not have reliable information concerning the parallax of $\beta$ Cyg A, and the analysis performed in Section \ref{sec:finalOrbit} resulted in an orbital  parallax barely significantly smaller than that of $\beta$ Cyg B. This uncertainty on the parallax translates into a large uncertainty on the line-of-sight separation between $\beta$ Cyg A and B, which is a key ingredient to determine if the system is gravitationally bound. For these reasons, since it is very difficult to obtain a complete and accurate orbital solution for the Albireo stellar system, we decided to study its dynamical nature via Monte Carlo simulations. Our simulations are loosely based on the combined analysis of RVs and astrometric data in \citet{pinamontietal2018}, following the technique from \citet{hausermarcy1999}.

We took into account the possible distances and proper motions for \bca\ derived from the final orbital solution we presented in Sect. \ref{sec:finalOrbit}. 
We adopted the mass value of $\beta$ Cyg B listed in Table \ref{Tab:PhysPar}, and the positions of the two components from \gdrtwo. The proper motion of $\beta$ Cyg B adopted was that of Brandt (Table \ref{tab:astrom_tbl}), while the parallax value adopted of \bcb\ was the weighted mean value derived in Sect. \ref{sec:absolute_ast} ($8.33 \pm 0.13$ mas).  The proper motion and mass for $\beta$ Cyg A, is obtained from the final orbit solution (Table \ref{tab:combined_full_model}).
Since the position in the sky of $\beta$ Cyg A depends on the orbital motion of the Aa with respect to the barycentre, we added in quadrature the semi-major axis of the orbit to the error on the position given by \gaia . We adopted the systemic radial velocity for \bca\ from Table \ref{tab:combined_full_model}, and the value $-18.8 \pm 2.20$ \kms from \citet{Kharchenko2007} for B. We then randomly generated all parameters for \bcb, as well as the position of \bca\ and it's radial velocity, with Gaussian distributions centered on the mean value of each parameter and standard deviation equal to their respective uncertainty. The other parameters for \bca\ (i.e. its parallax, mass, and proper motions) were instead randomly drawn from the posterior distributions obtained from the DE-MCMC analysis in Sect. \ref{sec:finalOrbit}, to preserve possible correlations. We generated a sample of $1\,000\,000$ realizations.

Taking into account the possible parallax distribution for $\beta$ Cyg A derived in Section \ref{sec:finalOrbit}, we can compute the line-of-sight separation $z$, and we thus have the complete set of possible relative positions and velocities, as the relative positions in the sky, ($x,y$), of B with respect to A, can be derived from \gaia\ astrometry, and the three components of the relative velocity $V_x,V_y,V_z$, from the proper motions and RVs.
For the system to be gravitationally bound, the total energy must be negative:
\begin{equation}
\label{eq:bound_energy}
E = {1 \over 2} {{M_A M_B} \over (M_A + M_B)} v^2 - {{G M_A M_B} \over r} < 0,
\end{equation}
where $r$ and $v$ are the relative positions and velocities of B with respect to A.

\begin{figure}
   \centering
   \includegraphics[width=.45\textwidth]{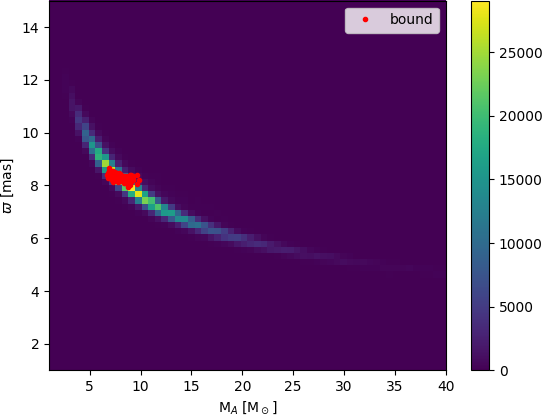}
      \caption{Probability density of possible \bca\ parallaxes $\pi_A$ and masses $M_\text{A}$ for the complete sample drawn from the posteriors of the DE-MCMC model from Sect. \ref{sec:finalOrbit}. The red dots mark the subset of solutions resulting in \bca\ and B being bound.}
         \label{fig:par_bound}
\end{figure}

We then study the distribution of parallaxes and masses of $\beta$ Cyg A for the bound systems, which are shown in Fig. \ref{fig:par_bound}, compared to the general distribution of parallaxes and masses resulting from the analysis in Sect. \ref{sec:finalOrbit}. It is worth noticing that the fraction of bound solutions is very small, with only $\sim 0.01 \%$ of the systems gravitationally bound, and that they are all concentrated in a very small island of the parameter space. The set of bound solutions always correspond to cases in which the parallax of $\beta$ Cyg A is very close to the value of $\beta$ Cyg B ($\varpi_{A,\text{bound}} = 8.32^{+0.11}_{-0.12}$ mas) and with a total mass ($M_{A,\text{bound}} = 7.80^{+0.89}_{-0.56}$ M$_\odot$) smaller than the best fit value for $M_\text{tot}$ listed in Table \ref{tab:combined_full_model}, 
and very close to the spectroscopic estimate of the total stellar mass of \bca\ determined in section \ref{sec:spectro}. However, the mass ratios of these bound systems is nevertheless found to be $q = 1.21^{+0.15}_{-0.14}$ , inconsistent with  of the mass ratio obtained from the same values in Table \ref{Tab:PhysPar}, but instead very close to the value obtained in Sect. \ref{sec:finalOrbit}.

As an additional test, we studied this best-case region of the parallax-mass parameter space that allows bound orbits, to better understand the influence of the other parameters on the dynamical state of the system, in particular of the relative velocity of the two components. For this test the mass value of \bca\ listed in Table \ref{Tab:PhysPar} is used, since it was very close to the value obtained above for the bound solutions, and we randomly generated its values with a Gaussian distribution, instead of drawing them from the posterior as previously done.

\begin{figure}
   \centering
   \includegraphics[width=.45\textwidth]{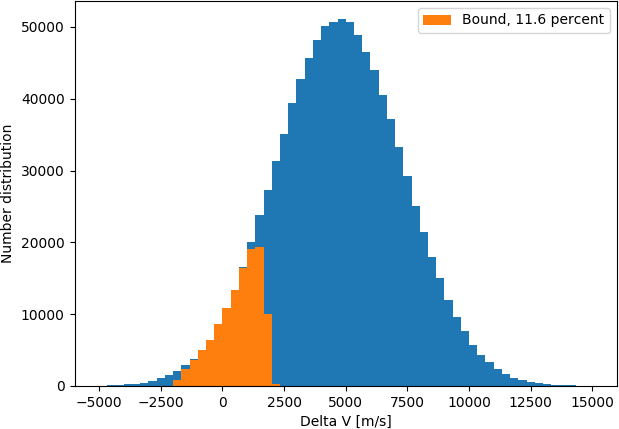}
      \caption{Distribution of possible relative line-of-sight velocities $V_z$ between \bca\ and B (blue) and the subset of these relative velocities that potentially allow bound orbits. }
         \label{fig:vz_bound}
\end{figure}

We then derived as before the relative positions in the sky, ($x,y$), and the three components of the relative velocity $V_x,V_y,V_z$. 
Without constraining the separation along the line-of-sight $z$, we can study the minimum energy of the system $E_\text{min}$, computed by substituting  $r = \sqrt{x^2 + y^2 +z^2}$ in Eq. \ref{eq:bound_energy} with $r_{xy} = \sqrt{x^2 + y^2}$. If $E_\text{min} >0$ there will be no possible value of $z$ for which the system is bound. Figure \ref{fig:vz_bound} shows the distribution of the relative line-of-sight velocity $V_z$ for the total sample and for the subsample of potentially bound systems where $E_\text{min} < 0$. In only $11.7\%$ of the realizations the system could be bound, and they all correspond to values of $V_z$ much closer to zero than the mean value: $V_{z,\text{bound}} = 0.80^{+0.71}_{-1.14}$ km s$^{-1}$. This suggests that, even in the best-case scenario in which the two components are nearly at the same distance, the observed difference in the relative line-of-sight velocity allows less than a 1 in 8 possibility that A and B are bound.

\subsection{Motion of the Albireo system}
\label{sec:baricenter}

With estimated mass for \bca\ and B, as well as their proper motion, we can infer the proper motion of the Albireo system, that is, of its barycentre.  If the two stars are bound their orbital velocities, with respect to the barycentre, are in opposite directions, and in consequence the components of these velocities tangent to our line-of-sight. In addition, the ratio of the magnitudes of these velocities are as the inverse ratio of their masses. These assumptions are also valid if A and B have recently become unbound, that is, their velocities with respect to their barycentre will be in opposite directions and their ratio as the inverse of their masses. If A and B are bound we can (and should) assume that they are at the same distance, so that the ratio of the projected orbital velocities also holds for their orbital proper motions, that is
\begin{equation}
\frac{\| \vec \mu_a \|}{\| \vec \mu_b \|} = \frac{M_B}{M_A} \equiv q_{AB} \, ,
\end{equation}
where $\vec \mu_a$ and $\vec \mu_b$ are the orbital proper motions of A and B with respect to their barycentre. These ratios will hold also for the unbound case, if the separation of A and B is negligible with respect to their distance. We can thus decompose the total proper motion into two components: the proper motion of the barycentre, $\vec \mu_s$, and the orbital proper motion from the velocity with respect to the barycentre. That is, the systemic proper motion of the \bca\ system is:
\begin{equation}
\vec \mu_A = \vec \mu_s + \vec \mu_a \, ,
\end{equation}
where $\vec \mu_a$ the component of this proper motion due to its velocity with respect to the barycentre of Albireo, and similarly for $\vec \mu_B$. Using the above ratio and solving for $\vec \mu_s$ we have
\begin{equation}
\vec \mu_s = \vec \mu_B + \frac{(\vec \mu_A - \vec \mu_B)}{(1 + q_{AB})} \, ,
\end{equation}
keeping in mind that the orbital proper motions are in opposite directions, that is, $\vec \mu_a = -q_{AB} \vec \mu_b$. Now for \bca\ we consider two possible masses. If bound and/or at the same distance as \bcb , $M_A = 7.9$M$_\odot$.  Otherwise we can assume $M_A = 9.47$M$_\odot$, as estimated in Section \ref{sec:finalOrbit}. In the first case we find the proper motion of the Albireo system to be $\vec \mu_s = (0.50, -0.57)$\masyr, while in the second case it is $\vec \mu_s = (0.58, -0.53)$\masyr.

\section{Kinematic analysis of the Albireo system: Relation between \bca\ and \bcb}
\label{sec:kinematic}

In the previous section we showed that, given our current observations, \bca\ and B are very unlikely to be a bound system. Or, if one prefers, that our current observations are unable to resolve the question of whether they are bound or not. In any case, the more relevant scientific question is not whether \bca\ and B are currently gravitationally bound or not, but whether they have a common history, age and origin. Under the assumption that they are at the same distance, section \ref{sec:evolution} shows that \bca\ and B are coeval, but in fact we do not know that A and B are at the same distance. In the following we will show that the question of having a common origin can be answered in the affirmative, by a probabilistic approach, and to a very high probability.

The century-old discussion whether Albireo\,A and\,B are physically connected or merely a coincidental optical double is based on their angular proximity on the sky. But, somewhat surprisingly, this argument apparently has never been elaborated quantitatively. By doing so, the question could have been largely settled already 23 years ago, in a probabilistic way.

\subsection{Two dimensions}
\label{twod}

The \hip\ Catalogue by definition includes a complete celestial census down to apparent magnitude V=7.9. Thus by construction it contains a complete census of all B~stars out to distances well beyond Albireo's: The absolute magnitude of an A0~star is M$_V$=+0.6, thus it reaches apparent magnitude V=7.9\footnote{Note that interstellar extinction is negligible within 120\,pc on the whole sky.} at 290\,pc.  

The New Reduction of the \hip\ Catalogue \citep{NewHipp2007} contains 
37696 stars with $\varpi>$7.0\,mas. Using the compilation by \citet{Anderson2012}, we find that among these there are 
591 B\,stars of all types, distributed almost evenly over the whole sky. For any given member of this set, the probability to find another unrelated member within 34.5\,arcsec is only 1.3\,10$^{-6}$ (590 times the solid angle covered by a circle of radius 34.5\,arcsec divided by the 4$\pi$\,radians of the whole sky). Thus already from this simple two-dimensional argument alone, the credibility of a physical connection between Albireo\,A and\,B is very high by all scientific standards.

\subsection{Three, five and six dimensions}

This low probability of an optical pair can even be further reduced by introducing additional observed quantities to the argument. 

Firstly, the distances of the two stars are roughly equal. Taking the least favourable combination of measured parallaxes for A and B, namely the \gaia\ DR2 ones, we can omit the inner part of the narrow spatial cone set by the positional agreement on the sky. This reduces the probability derived in Section \ref{twod} above by a mild factor of 0.6.

Secondly, and more importantly, the proper motions are roughly equal, too. 
The proper motion of \bcb\ is indisputably close to the \gaia\ DR2 value of (-1,-0.5) \masyr, to within 1 \masyr .  Above, we determined the systemic motion of Albireo\,A to be about (1.22$\pm$0.2,-0.17$\pm$0.6)\,\masyr (Table \ref{tab:combined_full_model}). 
At 120\,pc distance, the total proper-motion difference of about 2.3\,\masyr translates into a difference of tangential velocity by only 1.3\,\kms. 
This small difference must be compared with the velocity dispersion of young stars in the solar neighbourhood, which is slightly below 10\,\kms in all three spatial dimensions \citep{Binney2000}.

Assuming the dispersion of $\sigma_v=$10\,\kms, a conservative estimate of the probability of finding a coincidental partner to a given star within an interval $\pm\Delta v_t$ can be computed from the density function of a Gaussian distribution near its centre: $dP/dv=(\sqrt{2\pi}\sigma_v)^{-1}$. 
Applying this to both coordinates of the tangential velocity (using the uncertainty rather than the actual small value of the difference in declination), and multiplying with the 2-d probability found above, the total probability for an unrelated optical pair Albireo~AB is reduced to 1.3\,10$^{-8}$. 
Adding the less precise difference in radial velocity to the argument, the probability is lowered further, to 5\,10$^{-9}$. 

This means --- to a very high probability --- that the wide pair of the Albireo system, even if unbound, very likely has a common origin and is indeed coeval. 
 
%



\subsection{A possible Albireo moving group}
\label{sec:group}

A third --- less quantitative, but scientifically even more interesting --- way to argue for a physical connection between Albireo~A and~B would be the detection of an entire moving group or star cluster at the same distance and space velocity. We therefore conducted a careful search for possible additional members of the Albireo system. This search was in part motivated by the recent discovery of a formidable star cluster around another pair of B~stars, namely $\beta$ Lyrae \citep{Bastian2019}, the long-known existence of a similar cluster around the pair $\delta$ Lyrae, and of several moving groups around other B~stars.  
 
Drawing a tight 3-dimensional box around the \hip\ parallax of Albireo~B and the proper motion of the Albireo system, $\vec \mu_s = (0.50, -0.57)$\masyr (see Section \ref{sec:baricenter}), we found four more stars (in addition to Albireo~B) in a circle of 5\,degrees around the position of Albireo on the sky. Three of them lie within 0.8 degrees from that position. The additional members of the group are listed in Table\,\ref{tab:group}. Details on their selection are given in the Appendix.

\begin{table*}
\centering
\caption{The candidate members of the suspected Albireo moving group. The columns give the \gdrtwo\ star number, right ascension and declination, parallax, magnitude, colour and angular separation from Albireo. }
\label{tab:group}
\begin{tabular}{ccccccc}
\hline
SourceId \gdrtwo\ & alpha  & delta & parallax & G & BP-RP & rho \\ 
  & degrees& degrees & mas & mag & mag & degrees \\ 
\hline
 2026157766897183360 & 292.2059 & 28.1536 & 7.9 & 14.7 & 2.4 & 0.4 \\ 
 2026178417101258624 & 292.7363 & 28.3575 & 8.3 & 15.9 & 2.8 & 0.4 \\ 
 2026300016218003072 & 292.7387 & 28.7854 & 8.3 & 13.6 & 1.9 & 0.8 \\ 
 2032809846604646400 & 293.7991 & 31.2592 & 8.2 & 17.1 & 3.1 & 3.7 \\ 
\hline
\end{tabular}
\end{table*}

As explained in the Appendix, the expected number of stars in the box is only 0.017 stars within the 0.8~degree radius, and 0.37 stars within the 3.7~degree radius. So, we have a strong over-density, especially in the smaller circle on the sky. The Poissonian probabilities of finding three stars in the smaller and four stars in the larger circle are 9\,$10^{-7}$ and 5\,$10^{-4}$, respectively. Thus, with quite high probability we have identified a sparse, but spatially well concentrated moving group surrounding the long-known triple (now probably quadruple) Albireo system. This lends additional credibility to the physical connection of Albireo\,A and\,B. 

The reality of the moving group could be confirmed beyond any reasonable doubt by measuring the sixth phase space component, namely the radial velocity, of the four suspected members. A moderate precision of 1-2\,\kms would be sufficient. But due to the faintness of the stars, this task is outside the scope of the present study.  

In passing we note that the Jacobi limit (``tidal radius'') of a stellar system of 11.6\,M$_\odot$,  i.e.~having the sum of the masses of Albireo Aa+Ac+B derived from the spectroscopy in section \ref{sec:spectro} (Table \ref{Tab:PhysPar}), is about 3.0\,pc.  At 120\,pc distance this translates into 1.5 degrees. Members of a bound system are expected to be found mainly within two Jacobi radii from the center of mass, i.e.~within 3\,degrees in this case. This means that three out of our four member candidates are well within this limit, and the fourth one marginally outside.


\section{Evidence of hidden mass in \bca}
\label{sec:discussion}

As shown in section \ref{sec:spectro}, the properties of the known stellar components of the Albireo system are consistent with it being coeval, with an age of 99 Myr, 
under the hypothesis that all stars in the \bca\ system are at the same distance (120pc). This result is also consistent with the orbital solution derived from the radial 
velocity data and (relative) speckle astrometry (section \ref{sec:RVspeckle_orbit}) which, if assuming a distance of 120pc, results in a mass ratio for 
\bca a/Ac of $q=0.53_{-0.09}^{+0.13}$, consistent with the value of $q=0.52$ derived from the spectroscopic/photometric analysis under the same hypothesis. 
Only one detail seems out of place, namely the derived luminosity and temperature of Ac is not as expected for a star of 99 Myr.
This can be considered a minor inconsistency, given the large uncertainties in the derived stellar parameters of Ac, unavoidable given its proximity to Aa. 

This tidy picture is completely disrupted when we bring in the absolute astrometry, which allows us to derive an orbit solution completely consistent with the 
radial velocity and speckle data, but which gives us a mass ratio of $q=1.25^{+0.19}_{-0.17}$,  a total mass of $M_\mathrm{tot} = 9.47_{-3.24}^{+5.88}$ M$_\odot$ for the \bca\ system, 
and an orbital parallax that implies that the separation between Albireo A and B is about 10\,pc.
While the total mass agrees within the uncertainties with that determined from the spectrophotometry, the ratio of the masses is inconsistent with the individual masses estimated in Section \ref{sec:evolution} that suggest $q = 0.53$. 

Might there be problems with the absolute astrometry from either \hip\ or \gaia, or has our interpretation of the absolute astrometry gone astray?  

For \hip , the \bca\ system is unresolved and, not withstanding its brightness, should not present particular measurement problems. For \gaia, \bca\ is more problematic, 
given its brightness as well as the fact that, with a separation of $\approx 0.4$ arcsec, the system is marginally resolvable, if the fainter member is not completely 
lost in the glare of the primary. These effects might introduce important bias in both the \gaia\ measured proper motions and parallax, but should not introduce bias 
in the measured position. However, we stress that our result does not depend on the \gaia\ proper motion. Indeed, the mass ratio $q$ and systemic proper motion can 
be analytically solved for using only the \hip\ and Brandt proper motions, the latter depending only on the \gaia\ position, together with the relative proper motion 
and displacement of the secondary (Ac) with respect to the primary (Aa) at the \hip\ epoch, based on the fit to the speckle data.  One can show that
\begin{equation}
Q \equiv \frac{q}{1+q} = \frac{\| \vec{\mu}_B - \vec{\mu}_H \|}{\| \vec{\mu}_h - \frac{\Delta \vec{x}_\mathrm{rel}}{\Delta t_{HG}} \|} \, ,
\end{equation}
where $\Delta \vec{x}_\mathrm{rel}$ is the relative displacement vector of Ac between the \gaia\ and \hip\ epochs, and $\Delta t_{HG} = 24.25$yrs is the difference 
between the \hip\ and \gdrtwo\ epochs. From the relative orbit solution we have $\vec{\mu }_h = (15.31,10.72)$\masyr\ and $\Delta \vec{x}_\mathrm{rel} = (0.145,0.430)$\,arcsec,
 giving us $q=1.29$ and a systemic proper motion of $(1.33,-0.023)$\masyr, in complete agreement with the solution derived in section \ref{sec:finalOrbit}. 
 That the \gaia\ proper motion has little weight in the final orbit solution is a consequence of the larger (inflated) uncertainties assigned to these measurements. 

This leaves us with possible problems in our analysis, and in particular our assumptions. We have assumed that \hip\ and \gaia\ are both measuring the position of 
the primary, Aa, that is, the measured photocenter corresponds to the position of Aa. However, a more careful analysis, taking into account the flux ratio of the 
two components in the \hip\ and \gaia\ passbands, would only make $q$ even larger, exacerbating the problem of needing additional mass in the Ac component. 

We've also assumed that the speckle reference frame is the same as that of \hip\ and \gaia\ . However, we expect that at most there may be a rotation between the two frames no larger than $\delta \theta =0.1^\circ$, leading to corrections of the order of $\rho\sin \delta \theta =0.7$\,mas (where $\rho = 400$mas 
is the angular separation of Aa/Ac), which is much too small to possibly account for large systematic errors in the astrometry. 

Trusting our results, we conclude that the absolute astrometry from \hip\ and \gaia\ indicate that there is likely hidden mass in the \bca\ system. 

One obvious way to increase the mass of Ac without offending the observations is to hypothesize that it is itself a binary. The maximum mass for Ac would be an 
equal-mass binary with two stars of half the luminosity as observed for Ac, which would also explain its over-luminosity. At a distance of 120pc a star with half 
the luminosity of Ac would have a mass of 2.25$M_\odot$, bringing the total mass of \bca\ to 9.7$M_\odot$ and raising $q$ to 0.87. This would bring the total mass 
in agreement with the total mass determined using the absolute astrometry, but still outside the expected range for $q$, (1.25$^{+0.19}_{-0.17}$). 

If however \bca\ is more distant than B, as suggested by our final orbit solution, then both Aa and Ac will be more luminous.  The primary Aa would, consequently, 
be more massive but still coincide, within the uncertainties, with the age of B. At 133 pc distance, as indicated by \hip\ for example, we obtain a mass of Aa of 5.4 $M_{\odot}$ with an 
age of 90 Myrs. However, Ac would now fit on a main sequence evolution track of 3.0 $M_{\odot}$, but even further from the then relevant 90 Myr isochrone. If instead Ac were an equal mass binary its total mass would be $2 \times (2.5) = 5.0 M_\odot$, resulting in a total mass for \bca\ of 10.4$M_\odot$, again consistent with that determined from the astrometry, but still with $q < 1$.

Another, more exotic, solution to our missing mass problem is that the light from Ac comes from a star with a black-hole companion. This hypothesis might also explain 
the anomalous temperature/luminosity of Ac, which indicates that it is more evolved than expected from an 99 Myr isochrone. A search for corroborating evidence in 
NASA's HEASARC database does not reveal \bca\ as being a high-energy source, however such emission is only expected if mass transfer has created an accretion disk 
about the black hole, and indeed is absent in known systems with candidate stellar mass black holes \citep{Rivinius2020, Liu2019}. Assuming the total mass of 5.23 M$_\odot$ for Ac derived in Sec. \ref{sec:RVspeckle_orbit} is the sum of the 2.7 M$_\odot$ luminous component and a 2.53 M$_\odot$ dark companion, we used a standard injection and recovery method of synthetic signals (see e.g. \citealt{Mortier2012}) in the residuals of the speckle relative astrometry to place an upper limit to the maximum 
separation between the black hole and the stellar companion. We find that, for assumed system distances of 130 pc and 120 pc and averaging over all orbital elements, we would have detected (at the 95\% confidence level) clear periodicities in the residuals for orbital periods $>3.1$ yr and $>2.5$ yr respectively, corresponding to angular orbit sizes of $\gtrsim15$ mas (approximately twice the size of the rms of speckle astrometry residuals). 
Based on a Generalized Lomb Scargle periodogram analysis \citep{Zechmeister2009}, we see no evidence of periodic motion in the residuals of the speckle data, 
therefore the companion must have a significantly shorter orbital period. 


Evidence of an unseen high-mass companion to Ac would be confirmed if Ac's radial velocity significantly differed from that of Aa. A search of the literature 
for radial velocity measurements of Ac, whose Balmer series is clearly visible in the \bca\ spectrum, resulted in only one find: eight measures by E. Hendry, 
published with her measurements of the primary \citep{Hendry81}. These measurements are, not surprisingly, rather uncertain, but show a significant variance. 
However their mean is also significantly biased with respect to the systemic radial velocity of \bca , bringing into question their reliability.  
Given that no others in the field have dared to measure (or publish) is a good indication of the challenge of measuring Ac's radial velocity, but if indeed Ac 
has a black hole as a companion, high precision might not be needed if the orbital inclination is not small. 

We note that there are also LWP and SWP high-resolution spectra from the 1980's in the 
IUE archive, which should be dominated by the flux from the hotter Ac component. 
We did not consider using this potential radial velocity 
information as the read-out of the reticon cameras of IUE
was rather noisy: Even in well exposed spectra,
signal-to-noise rarely exceeds 20. More importantly the spectral resolving power of the
high-resolution mode is also very modest by today's standards,
varying between 12,000 and 15,000. This results in a measurement error of up to 0.07\AA\ 
(for details, see \citet{Boggess1978}
), or 7 to 10\kms, much larger than the amplitude of the orbital velocity 
expected from the Aa/Ac pair alone. However, given the possibility that orbital motion with a close unseen companion might be detected, it may be worth reconsidering these measurements. 

In any case, our determination of the total mass of the \bca\ system based on the astrometry and radial velocity data indicates that there is likely a fourth, 
hidden component, making Albireo a hierarchical quadruple system.

\section{Conclusions}
\label{sec:conclusions}


We have here presented the first well-constrained orbit solution for the close binary comprising the primary of the wide pair of the Albireo system, \bca , 
finding a period of $121.65^{+3.34}_{-2.90}$ yr and low eccentricity.  This is mainly accomplished by combining the speckle data, providing relative astrometry 
over about one fourth of it's orbit, with radial velocity measurements spanning a baseline of more than 120 years. 

With the addition of absolute astrometry from \hip\ and \gaia , we are also able to estimate the systemic proper motion of \bca\ and the mass ratio of the Aa/Ac pair, 
as well as the total mass of the system. Our final orbit solution also gives an orbital parallax estimate. 
Our determination of the systemic motion of \bca , (1.22$\pm$0.2,-0.17$\pm$0.6)\,\masyr, 
is in good agreement with the existing long-term measurements of the proper motion in e.g.~FK4 and FK5\footnote{These two values, despite being derived from observations 
over more than a century, will still contain some portion of the averaged relative motion due to the 120-year orbit.}, and our estimated orbital parallax for \bca , 
7.75$\pm$1.2\,mas, is in good agreement with the New \hip\ measure of 7.51 $\pm$0.33\,mas. These parallax measurements for \bca\ suggest that the line-of-sight distance between 
the A and B components of Albireo may be about 10 to 20 pc, with the implication that A and B are very likely unbound. Nevertheless, such a separation is not 
in conflict with the hypothesis that these stars were born from the same birth cluster: Consider a common formation of the triple, but in a kinematic configuration 
that was unbound from the very beginning. The velocity spread in unbound star-forming complexes is in the range of 0.5~to 1\,\kms typically. An initial velocity 
difference of only 0.2\kms would have separated the two partners by a full 20\,pc over the past 100 million years. 

On the other hand, our final orbit solution for the \bca a/Ac system does not exclude the possibility that A and B are indeed bound, though our current measurements 
make this a very unlikely possibility. Clarifying the dynamical state of this system will require not only more precise astrometry, but better radial velocity determinations 
for both A and B. 

Regardless of whether the A and B components are bound, we find there is a very low probability of Albireo A and B being a chance alignment of unrelated stars. 
This purely statistical argument based on the kinematics and positions of A and B is strengthened by the likelyhood that the stars are coeval (Sec.\,\ref{sec:evolution}), and the discovery of a few more stars that seem to be associated with the wide pair (Sec.\,\ref{sec:group}). Altogether, this indicates that the Albireo system is the residual massive core of a sparse dissolved --- or still dissolving --- star cluster.

The most interesting finding of the present study is the surprising fact that Albireo Ac is significantly more massive than Albireo Aa. This is in stark contrast to the 
photometric and spectroscopic analysis --- if Ac is considered to be a single star. It should be pointed out that our finding is very robust: It is independent of the 
assumed parallax. It is also independent of the orbit solution; it can be directly deduced from a comparison of the relative motion of Ac with respect to Aa, as traced 
by the speckle data (Figs. \ref{fig:PAplot} and \ref{fig:Sep_plot}), with the \gaia\ and \hip\ absolute proper motions of Aa. It is even independent of the \gdrtwo\ proper motion, as it clearly stands 
out from the \hip\ and Brandt proper motions alone (as described in Section\,\ref{sec:discussion}). 

Based on the data at our disposal, the additional mass of Ac is best interpreted in terms of the presence of a companion, suggesting that \bca\ may in fact be a triple system. 
If we trust the orbital parallax estimate (corroborated by its agreement with the \hip -based direct measurement), and our measurement of the mass ratio of Ac/Aa, it is difficult to reconcile the photometric and 
spectroscopic data available with the presence of a second luminous, massive component. This raises the intriguing possibility that Ac might be orbiting a dormant, stellar-mass 
black hole. The nature of the additional mass in Albireo Ac cannot be conclusively determined for the time being, but promising paths towards clarification do exist: these 
include targeting Albireo Ac with high-contrast imaging observations, adaptive-optics spectroscopy, radial velocity measurements, and very high-resolution interferometric astrometry. 

The foundation of these central findings consists of three parts: the robust new orbital solution including the measured radial velocities and relative astrometry from 
speckle data, the careful spectroscopic confirmation of the physical parameters of the observable stars, and the inclusion of the absolute astrometry from \hip\ and \gaia .
 The surprising mass ratio of Ac/Aa had already been found by \citet{bastian18}, who could not resolve the mystery at the time because they neither had the correct orbit, 
 nor the new astrophysical information.

We can make a number of predictions for \gdrthree\ which --- if they hold --- will support our findings. They are: the parallax of A and B should become more consistent 
than seen in DR2, the parallax of A becoming less problematic, and the presently discrepant proper motion in declination should be corrected for Aa, becoming more consistent
 with that predicted by our final orbit solution, and the kinematic and spatial proximity of additional faint partners should be sharpened and confirmed. The increased 
 precision and reliability of the \gaia \, EDR3 proper motions with respect to DR2 will allow them to significantly improve and confirm the anomalous mass ratio of the stars 
 in the \bca\ system, as well as the total mass and orbital parallax of the system.

\section*{Acknowledgements}

A special thanks to Benoit Famaey for sharing CORAVEL data, to Marco Scardia for sharing the latest PISCO measurement and useful discussions, to Elaine Hendry for clarifications on her observations, and to Rainer Anton for catching an error in the computation of one of the derived parameters in an earlier draft of the manuscript.  This work has made use of data from the European Space Agency (ESA) mission \gaia\ (\url{https://www.cosmos.esa.int/gaia}), processed by the \gaia\
Data Processing and Analysis Consortium (DPAC, \url{https://www.cosmos.esa.int/web/gaia/dpac/consortium}). Funding for the DPAC has been provided by national institutions, in particular the institutions participating in the \gaia\ Multilateral Agreement. KS acknowledges support from the CONACyT-DFG bilateral project No. 278156. MP acknowledges financial support from the ASI-INAF agreement n. 2018-16-HH.0. 

\section*{Data availability}

All data used in this paper is presented in tabular format in section \ref{sec:data}.  The spectra for which the HEROS radial velocities are based are available from the authors on request.  



\bibliographystyle{mnras}
\bibliography{biblio} 




\appendix

\section{Supplementary figures}

\subsection{Posterior distributions of orbital solutions}

\begin{figure*}
   \centering
   \includegraphics[width=.95\textwidth]{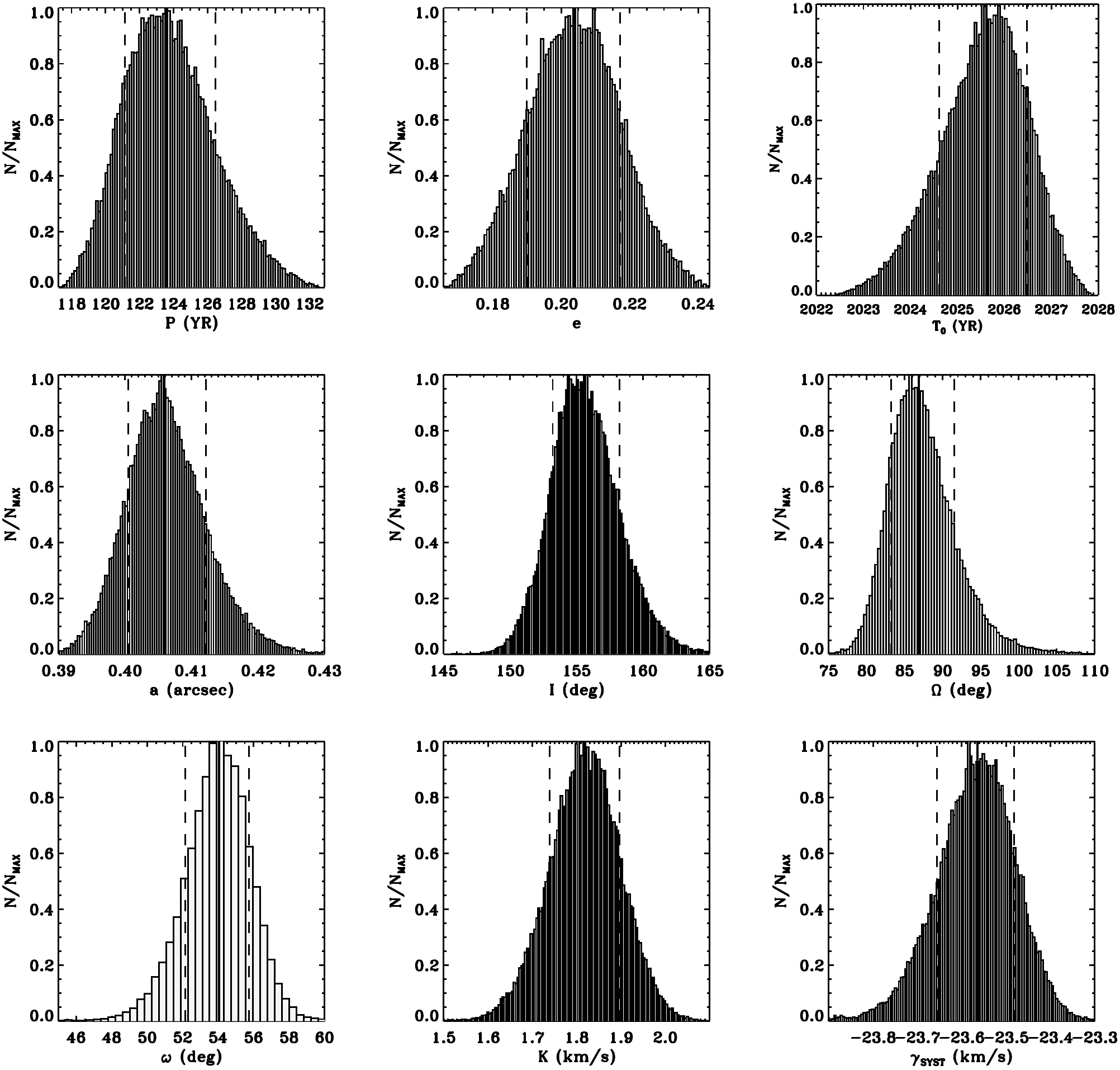}
      \caption{Posterior distributions of the fitted and derived parameters of the Keplerian model applied to the combination of the RV time series and speckle imaging astrometry (Table \ref{tab:combined_model}). 
      The vertical dashed lines denote the 16 th and 84 th percentiles, while the vertical solid lines identify the median values.}
         \label{fig:combined_posteriors}
\end{figure*}

\begin{figure*}
   \centering
   \includegraphics[width=.95\textwidth]{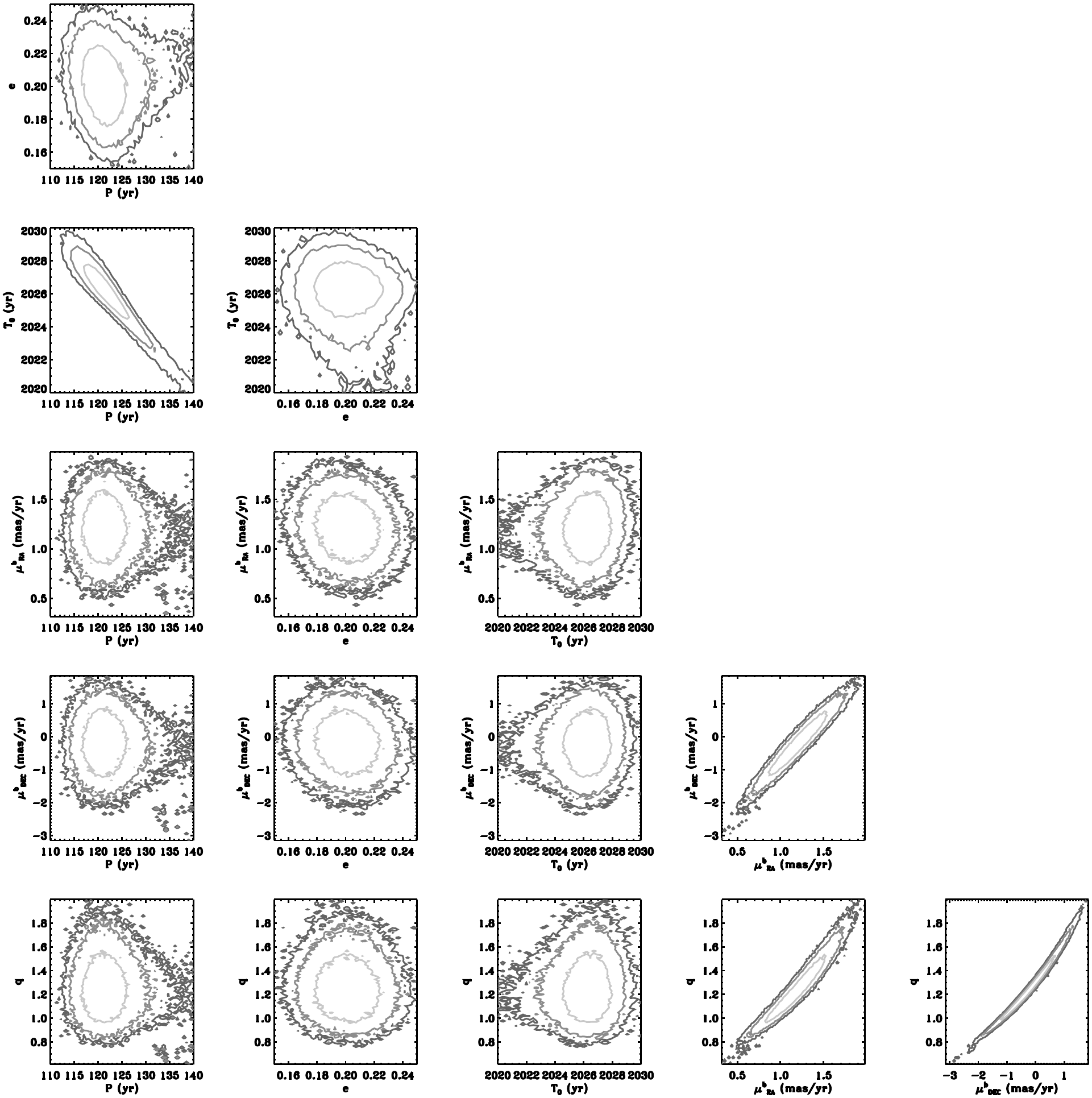} \\
      \caption{Joint posterior distributions for the model parameters explored in our DE-MCMC analysis (Table \ref{tab:combined_full_model}). Light grey contours indicate $1-\sigma$ ranges, while grey and dark 
      grey contours indicate $2-\sigma$ and $3-\sigma$ ranges, respectively.}
         \label{fig:joint_posteriors_full_model}
\end{figure*}

\begin{figure*}
   \centering
   \includegraphics[width=.95\textwidth]{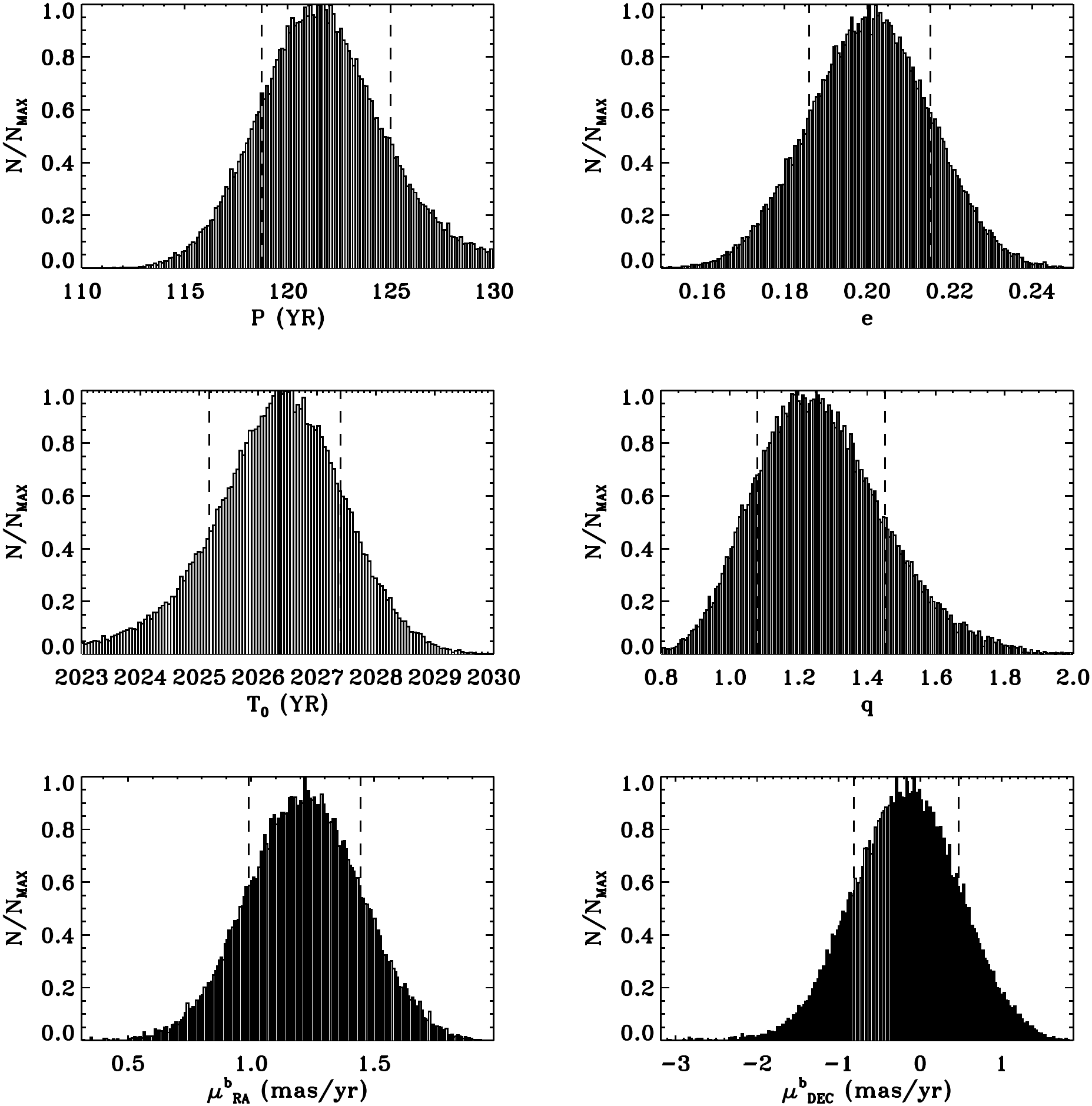} \\
      \caption{Posterior distributions of the four jump parameters of the Keplerian model for $\beta$ Cyg A based on the combination of the RV time series, speckle imaging 
      and absolute astrometry from \hip\ and \gaia\ (Table \ref{tab:combined_full_model}).  Same line coding as in Figure \ref{fig:combined_posteriors}.}
         \label{fig:jump_parameters_posteriors_full_model}
\end{figure*}

Figure \ref{fig:combined_posteriors} the posterior distribution of orbital parameters using both the RV time series and speckle imaging astrometry. Figures \ref{fig:joint_posteriors_full_model} and \ref{fig:jump_parameters_posteriors_full_model} the posterior distributions of the full joint solution using RV, speckle astronometry and the absolute proper motions. 

\begin{figure*}
   \centering
   \includegraphics[width=.95\textwidth]{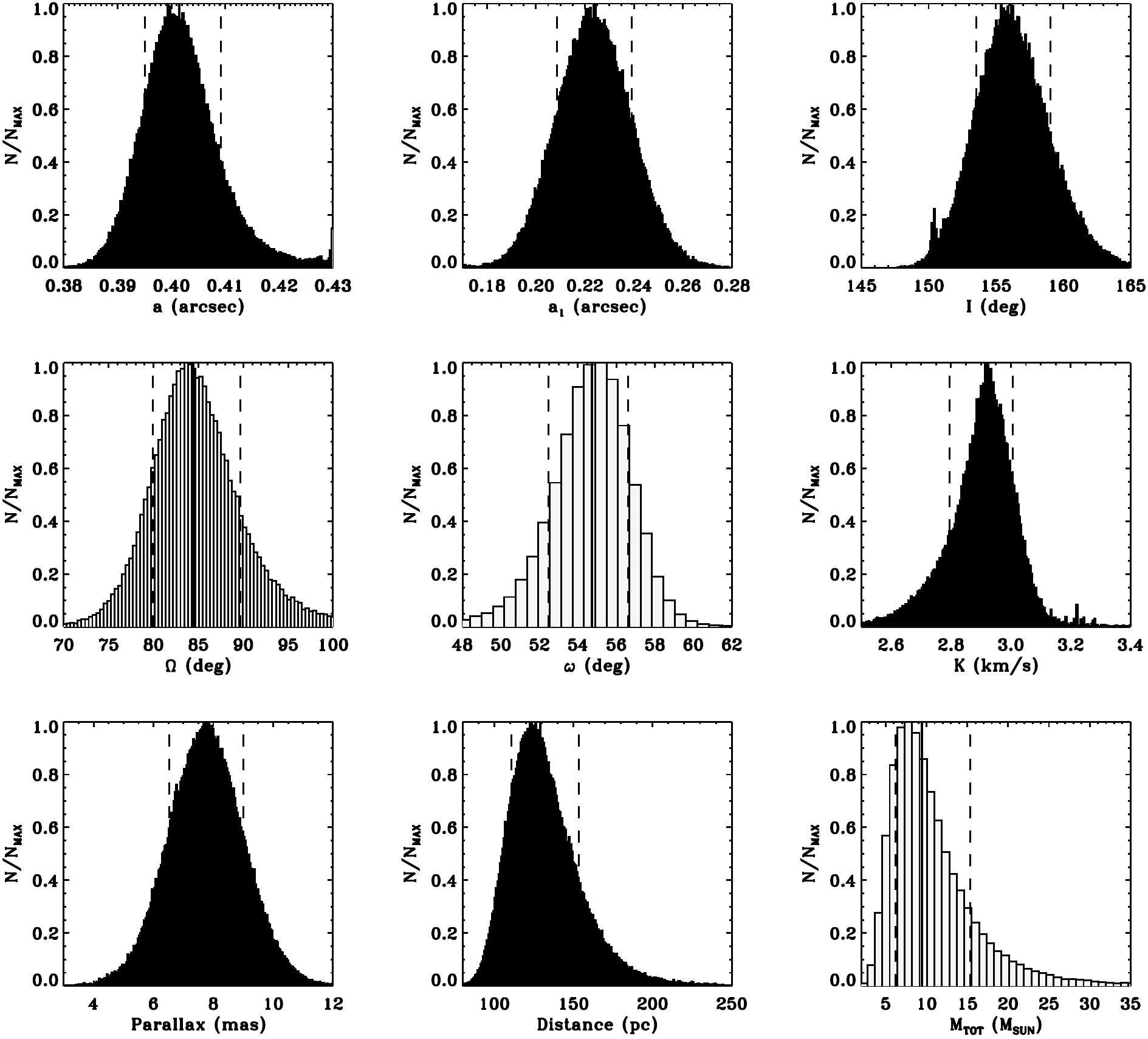} \\
      \caption{Posterior distributions of the derived parameters of the Keplerian model, parallax, and total system mass for $\beta$ Cyg A based on the 
      combination of the RV time series, speckle imaging and absolute proper motions from \hip\ and \gaia. Same line coding as in Figure \ref{fig:combined_posteriors}.}
         \label{fig:derived_parameters_posteriors_full_model}
\end{figure*}

\subsection{The high-resolution spectrum; two full pages}

\begin{figure*}
\includegraphics[width=0.96\textwidth]{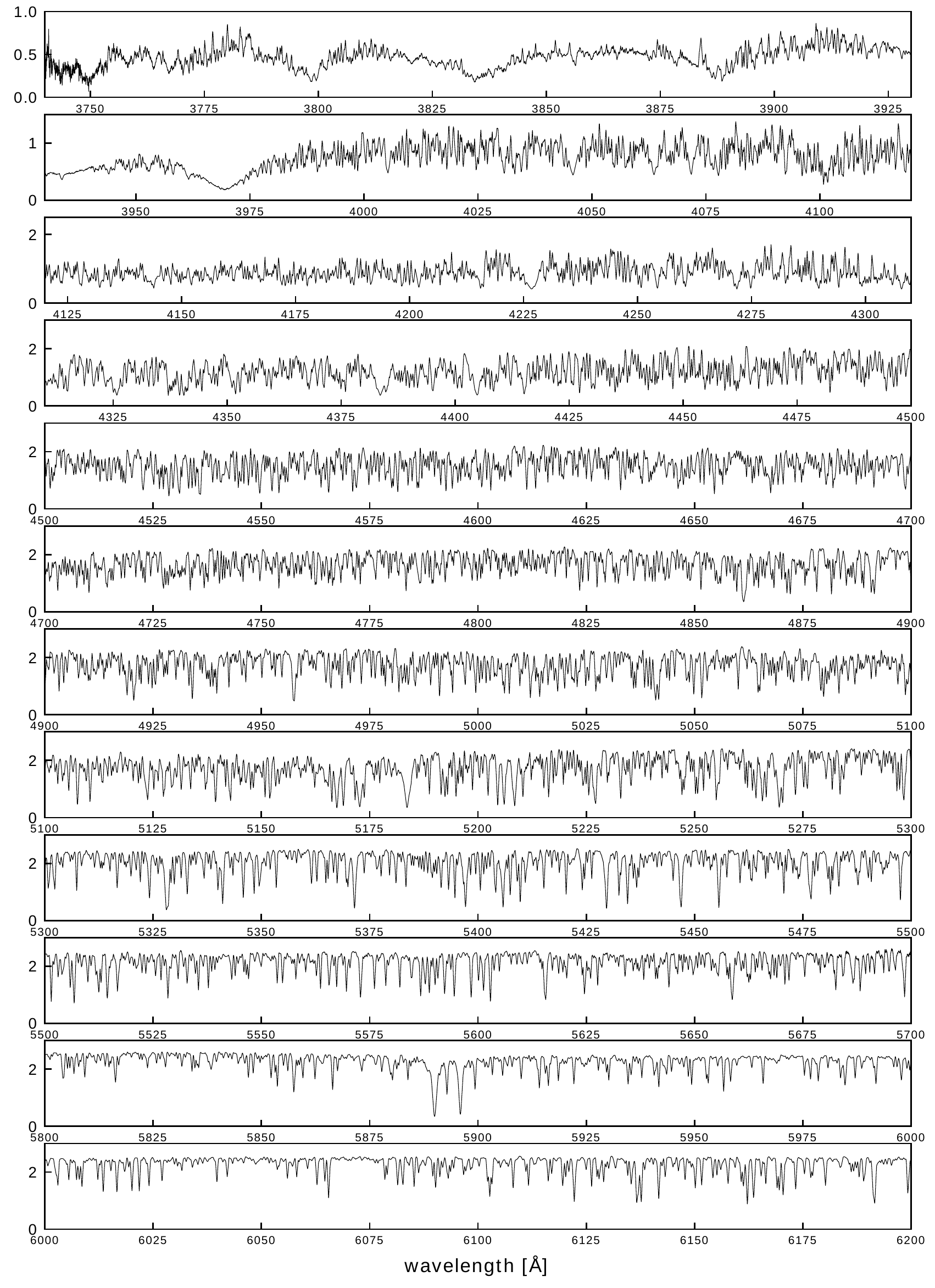}
\caption{Blue part of the high resolution composite spectrum of \bca a and \bca c.}
\label{fig_spec}
\end{figure*}

\begin{figure*}
\includegraphics[width=0.96\textwidth]{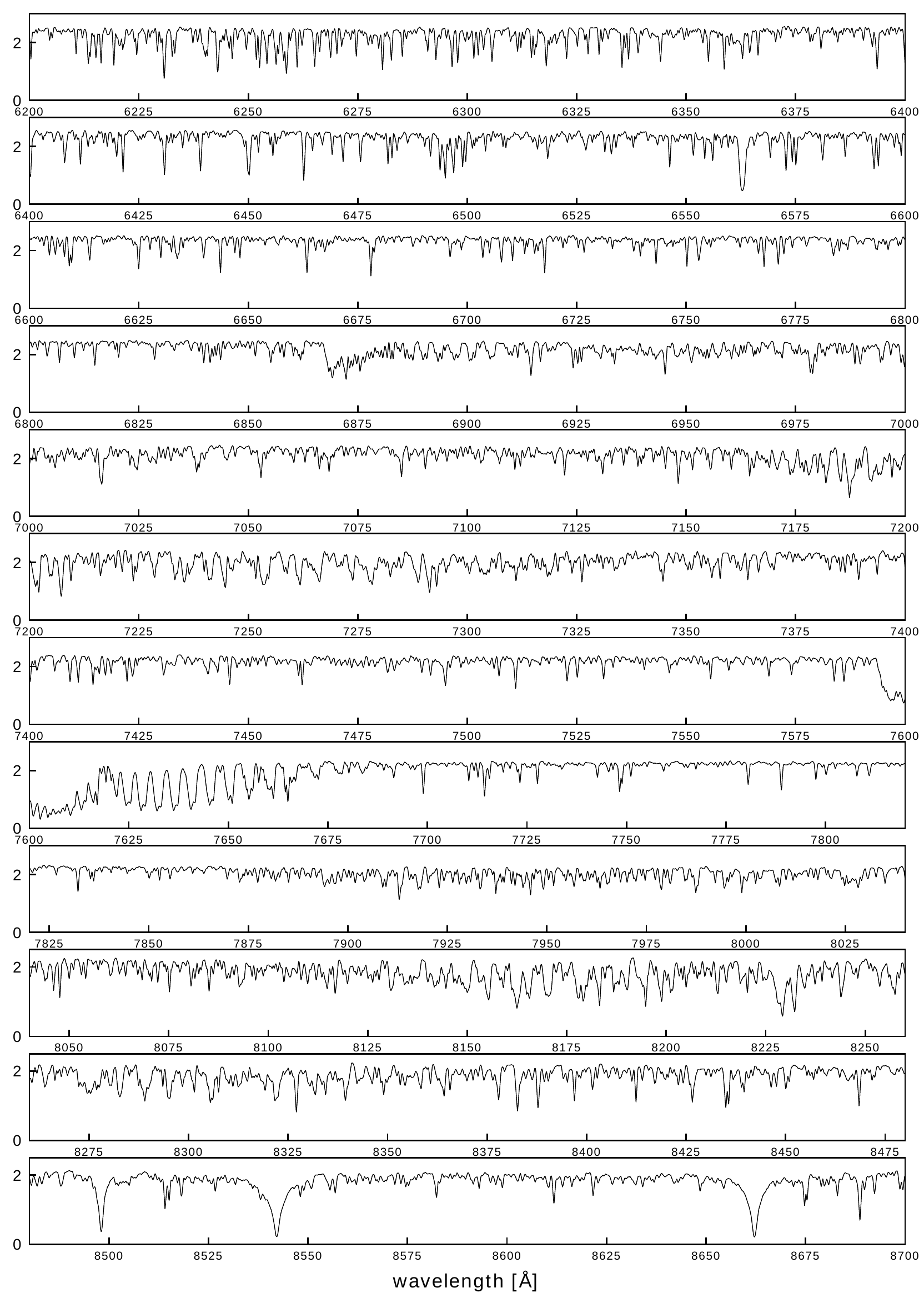}
\caption{Red part of the high resolution composite spectrum of \bca a and \bca c.}
\label{fig_spec2}
\end{figure*}

We present in Figures~\ref{fig_spec} and \ref{fig_spec2} the high resolution 
($R\approx 20,000$) spectrum obtained with the TIGRE telescope.

\section{Selection and mathematics of the moving group}

In order to do a five-dimensional search for stars possibly connected with Albireo kinematically, the search space should be centred on the most trustworthy spatial location (i.e.~in particular parallax) and space velocity of the Albireo triple system. For this purpose we use the \hip\ (2007) parallax and proper motion of Albireo~B. For the search intervals we chose $\pm$0.3\,mas in parallax (corresponding to the distance range from about 117.5\,pc to 126\,pc) and $\pm$1\kms in the two velocity coordinates (corresponding to 1.7\masyr at 122\,pc). We restricted the search to \gaia\,DR2 stars with $G<19$ (i.e.\,to $M_G<13.5$), so that the DR2 uncertainties in the measured parallaxes and proper motions are smaller than the search tolerances.
 
In this way we found the four stars listed in Table\,\ref{tab:group}. To estimate whether these can be chance field stars, the actually found number of four stars is to be compared with the expected number from the general galactic field. The following considerations define a conservative estimate.          

The expected number of stars to be found in a given 5-dimensional phase space volume of the local galactic disk simply is:
\begin{equation}
dN = \frac{dN}{dV} \Delta V \frac{dN}{dv_{t\alpha}} \Delta v_{t\alpha} \frac{dN}{dv_{t\delta}} \Delta v_{t\delta}
\end{equation}
where:
\begin{itemize}
\item $\frac{dN}{dV}$ is the total local number density of stars
\item $\Delta V$ is the 3-dimensional spatial volume under consideration
\item $\Delta v_{t\alpha}, \Delta v_{t\alpha}$ are the intervals under consideration in tangential
\item velocity $v_t$ in the directions of right ascension $\alpha$ and declination $\delta$, respectively.  
\item $\frac{dN}{dv_{t\alpha}}$ and $\frac{dN}{dv_{t\delta}}$ are the normalised densities in these velocity
\item coordinates, at the specific velocities under consideration.
\end{itemize}

The canonical value for $\frac{dN}{dV}$ can be taken from the Catalogue of Nearby Stars (CNS4). It is 0.12/(pc)$^3$. The spatial volume $\Delta V$, defined by the above parallax tolerance and a circular field on the sky around Albireo with angular radius~$r$, can be computed as follows:
\begin{equation}
\Delta V = 4\pi D^2 \Delta D * \Omega/(4\pi)
\end{equation}
where $D$ is the mean distance of the volume (120\,pc), $\Delta D$ is the adopted distance range (8.5\,pc, see above) and $\Omega= \pi r^2$ is the solid angle subtended by the sky field under consideration.

For the normalised densities in the velocity coordinates we conservatively assume that the space velocity of Albireo\,B is close to the mean galactic disk rotation. Thus we can use the normalised density close to the center of the local velocity ellipsoid, which for a Gaussian distribution of dispersion $\sigma$ is $(\sqrt{2\pi}\sigma)^{-1}$. Adopting --- again very conservatively --- a value of $\sigma$=10\kms (which is true only for very young stars), we find 0.017 and 0.37 as the expected number of stars in circular sky fields of 0.8\,degrees and 3.7\,degrees radius around Albireo. 
Simple Poisson statistics then yield the probabilities given in Section\,\ref{sec:dynamical} for finding three and four stars, respectively within these circular fields. 

No further stars were found out to five degrees radius. Also, no further ones were found in a four times larger search box in $v_{t\alpha}$, $v_{t\delta}$ space. The above conservative expected number of stars in this larger box would be about 0.07, 1.5 and 2.7 for $r$=0.8\,degrees, 3.7\,degrees and 5\,degrees, respectively. The fact that indeed no further stars are found in the larger phase space volume lends further credibility to the reality of the moving group and confirms the conservative nature of the above calculations.


\bsp	
\label{lastpage}

\end{document}